\documentclass[12pt,english]{article}
\usepackage[english]{babel}
\usepackage{amsmath,amssymb,amscd,color}
\usepackage{amsfonts}
\usepackage{mathrsfs}
\usepackage{mathtools}
\usepackage{graphicx}
\usepackage{url}
\usepackage{hyperref}
\usepackage{cite}
\usepackage{physics}
\usepackage{bbold}
\usepackage{braket}
\usepackage{verbatim}
\usepackage[font=footnotesize,labelfont=bf,width=.9\textwidth]{caption}
\usepackage{multirow}
\usepackage{boldline}
\usepackage{enumitem}
\allowdisplaybreaks

\hypersetup{
    colorlinks,%
    citecolor=blue,%
    filecolor=blue,%
    linkcolor=blue,%
    urlcolor=blue
}

\usepackage[textsize=scriptsize,textwidth=2.5cm]{todonotes}

\def\Ket#1{\left|#1\right\rangle}

\makeatletter
\renewcommand\section{\@startsection {section}{1}{\z@}%
                                   {-5.5ex \@plus -1ex \@minus -.2ex}
                                   {2.3ex \@plus.2ex}%
                                   {\normalfont\large\bfseries}}
\renewcommand\subsection{\@startsection{subsection}{2}{\z@}%
                                     {-3.25ex\@plus -1ex \@minus -.2ex}%
                                     {1.5ex \@plus .2ex}%
                                     {\normalfont\bfseries}}

\numberwithin{equation}{section}

\makeatother

\newcommand{\bea}{\begin{eqnarray}}
\newcommand{\eea}{\end{eqnarray}}
\newcommand{\be}{\begin{equation}}
\newcommand{\ee}{\end{equation}}

\newcommand{\eq}[1]{\begin{align}#1\end{align}}
\newcommand{\eqst}[1]{\begin{align*}#1\end{align*}}
\newcommand{\eqsp}[1]{\begin{equation}\begin{split}#1\end{split}\end{equation}}

\newcommand{\Z}{{\mathbb Z}}

\def\vp{\varphi}
\def\p{\partial}
\def\Vev#1{\big\langle#1\big\rangle}

\newcommand{\cC}{{\cal C }}
                
\newcommand{\cO}{{\cal O }}

\newcommand{\cN}{{\cal N }}

\newcommand{\vac}{|0\rangle}

\newcommand{\ie}{{\it i.e.~}}
\newcommand{\etc}{{\it etc.~}}

\newcommand{\zb}{{\bar z}}
\newcommand{\xb}{{\bar x}}

\renewcommand{\title}[1]{\vbox{\center\LARGE{#1}}\vspace{5mm}}
\renewcommand{\author}[1]{\vbox{\center#1}\vspace{5mm}}
\newcommand{\address}[1]{\vbox{\center\footnotesize\em#1}}
\newcommand{\email}[1]{\vbox{\center\footnotesize\tt#1}\vspace{5mm}}

\begin{document}

\begin{titlepage}

 \begin{flushright}

\end{flushright}

\begin{center}

\hfill \\
\hfill \\
\vskip 1cm

\title{Deforming Symmetric Product Orbifolds:\\ A tale of moduli and higher spin currents}

\author{Luis Apolo$^{a,b}$, Alexandre Belin$^{c,d}$, Suzanne Bintanja$^{a}$, \\Alejandra Castro$^{a}$, and Christoph A. Keller$^e$
}

\address{
${}^a$ Institute for Theoretical Physics, University of Amsterdam, Science Park 904, \\
1090 GL Amsterdam, The Netherlands
\\
${}^b$Beijing Institute of Mathematical Sciences and Applications, Beijing 101408, China
\\
${}^c$CERN, Theory Division, 1 Esplanade des Particules, Geneva 23, CH-1211, Switzerland
\\
${}^d$Department of Theoretical Physics, University of Geneva, \\ 24 quai Ernest-Ansermet,
1211 Geneva 4, Switzerland
\\
${}^e$Department of Mathematics, University of Arizona, Tuscon, AZ 85721-0089, USA}

\email{l.a.apolo@uva.nl, a.belin@cern.ch, s.bintanja@uva.nl \\ a.castro@uva.nl, cakeller@math.arizona.edu}

\end{center}

\vfill

\abstract{We analyze how deforming symmetric product orbifolds of two-dimensional \linebreak $\mathcal{N}=2$  conformal field theories by an exactly marginal operator lifts higher spin currents present at the orbifold point. We find on the one hand that these currents are universally lifted regardless of the underlying CFT. On the other hand the details of the lifting are surprisingly non-universal, with dependence on the central charge of the underlying CFT and the specific marginal operator in use. In the context of the AdS/CFT correspondence, our results illustrate the mechanism by which the stringy spectrum turns into a supergravity spectrum when moving through the moduli space. They also provide further evidence that symmetric product orbifolds of $\mathcal{N}=2$ minimal models are holographic.}

\vfill

\end{titlepage}

\eject

\tableofcontents
\newpage
\section{Introduction}

Symmetric product orbifolds of two-dimensional CFTs play a prominent role in holography. They give specific realizations of the AdS$_3$/CFT$_2$ correspondence arising from string theory \cite{Maldacena:1997re,Dijkgraaf:1998gf,Giveon:1998ns,Seiberg:1999xz,Argurio:2000tb,David:2002wn}: a renowned example is the appearance of the symmetric product orbifold of $\mathbb{T}^4$ as the weakly coupled CFT dual to tensionless strings on AdS$_3$ \cite{Gaberdiel:2014cha,Giribet:2018ada,Eberhardt:2018ouy}. More generally, symmetric product orbifolds naturally incorporate many features that are compatible with the large-$N$ behaviour that is expected in holographic setups \cite{Pakman:2009zz,Pakman:2009ab,Keller:2011xi,Belin:2014fna,Haehl:2014yla,Belin:2015hwa}. This makes them a good laboratory to explore aspects of AdS/CFT with control and accuracy.  

While a lot is known about these theories at the orbifold point, less is known about their behaviour under deformations, most explicit analyses being based on the aforementioned case of $\mathbb{T}^4$ \cite{Avery:2010er,Gaberdiel:2015uca,Keller:2019yrr,Guo:2020gxm,Benjamin:2021zkn}. Such deformations are needed to introduce interactions in the system, which hopefully drive the theory to a strongly-coupled point in the CFT moduli space. There one expects to find features of the holographic duality in terms of a semi-classical and local theory of gravity. In this work we explore the effects of deformations using conformal perturbation theory in a manner that is systematic and robust. Our aim is to establish which features of the deformations are universal for any symmetric product orbifold CFT, and which ones depend on the CFT that is being orbifolded.

A two-dimensional symmetric product orbifold CFT is constructed as follows: First we take a seed CFT$_2$ of central charge $c$ that we denote by $\mathcal{C}$. The symmetric product orbifold of $\cC$ is then defined as
\be\label{eq:sym}
{\rm Sym}^N({\cal C})\coloneqq \frac{\mathcal{C}^{\otimes N}}{S_N} \,.
\ee
Here $S_N$ is the symmetric group, which instructs us to identify the permutations of the seed theory. This procedure has an effect analogous to gauging the theory: we keep states that are invariant under $S_N$, the untwisted sector, and add new states labelled by conjugacy classes of $S_N$, the twisted sectors. An important and appealing aspect of the construction is that the spectrum of ${\rm Sym}^N({\cal C})$ is under analytic control and, moreover, is determined in an elegant way by the data of the seed $\mathcal{C}$. In this work we will focus exclusively on theories with at least $\mathcal{N}=(2,2)$ superconformal symmetry.

Because of the $\mathcal{N}=(2,2)$ supersymmetry, the symmetric product orbifolds can contain exactly marginal operators, which we can use to deform the theory away from the orbifold point. In other words, these theories can have a moduli space in which we can move around using exactly marginal operators that preserve supersymmetry. Here we will focus on marginal operators in the twisted sector: this assures that the deformation will break the orbifold structure in \eqref{eq:sym}, and thus introduce interactions among the copies of the product theory. Deforming \eqref{eq:sym} by such operators can be interpreted as turning on marginal gauge couplings, namely
\be \label{deformationintro}
{\rm Sym}^N({\cal C})(\lambda) = {\rm Sym}^N({\cal C}) + \lambda \int d^2z  \ \Phi(z,\bar{z}) \,.
\ee
Note that the deformed theory ${\rm Sym}^N({\cal C})(\lambda)$ is again a conformal field theory which admits a 't Hooft-like limit for an appropriate $N$ scaling of the coupling $\lambda$. 

It was shown in \cite{Benjamin:2015vkc,Belin:2020nmp} that requiring the existence of a twisted modulus already imposes a stringent condition on the seed $\mathcal{C}$: its central charge must be in the range $1\leq c\leq6$. In our work we explain in further detail which twisted moduli are allowed for a given central charge, and describe their composition. We will use this, combined with other requirements described in Sec.\,\ref{sec:subintro1}, to give precise results on the effect of turning on twisted moduli in symmetric product orbifold theories. 

The specific question we investigate here is the following. At the orbifold point, the symmetry of the CFT is much bigger than just the $\cN=2$ superconformal algebra. This comes from the fact that we have $N$ copies of this algebra, and any permutation-invariant combination of its currents will form a symmetry current of the theory, possibly of higher spin. Since this phenomenon arises from the specific form of the orbifold theory, we do not expect these currents to remain conserved as we deform the theory. That is, we expect them to be lifted under deformations, so that they acquire an anti-holomorphic weight and are therefore no longer chiral. The lifting of the currents under the deformation is also expected from holography, as we will explain below.

In this paper we will check explicitly if and how the symmetry currents that arise from the orbifolding procedure are lifted, with a  special emphasis on the theories proposed in  \cite{Belin:2020nmp}.  Our analysis will corroborate the expectation that these conserved currents are lifted in general. However, a surprising effect is that the magnitude of the lift is not universal: it depends non-trivially on details of the seed ${\cal C}$ and is sensitive to the effects of the orbifolding.

\subsection{Holographic CFTs and their ties to symmetric product orbifolds }\label{sec:subintro1}

Let us describe a further motivation for our work coming from holography. Holographic CFTs are theories that admit a dual emergent description in terms of semi-classical general relativity with weakly-coupled matter through the AdS/CFT correspondence \cite{Maldacena:1997re}. These requirements, in addition to the usual conditions such as unitarity, locality, and crossing symmetry, highly constrain the theory. From our current understanding of AdS/CFT, holographic theories need to satisfy the following properties. First, they must admit a large number of local degrees of freedom, meaning that they must be large-$N$ theories. Second, the dimension of the lightest operator with spin greater than two must be parametrically larger than one \cite{Heemskerk:2009pn,Afkhami-Jeddi:2016ntf, Meltzer:2017rtf, Belin:2019mnx, Kologlu:2019bco}. The size of the gap in dimension, often referred to as $\Delta_{\textrm{gap}}$, controls the higher derivative couplings in the gravitational theory. Third, while the CFTs must have a large number of local degrees of freedom, they must have only few operators of low scaling dimension. In other words, they must have a sufficiently sparse spectrum of low-dimension operators \cite{Hartman:2014oaa,Belin:2016yll,Belin:2016dcu}. Not surprisingly, it remains a formidable task to classify CFTs that satisfy these conditions.

Nevertheless, progress has been made towards classifying theories that meet some of the aforementioned conditions. One approach is to identify partition functions compatible with the holographic conditions. This is particularly powerful in the example of compact two-dimensional CFTs with at least $\mathcal{N}=(2,2)$ supersymmetry, whose  elliptic genus is a weak Jacobi form \cite{Kawai:1993jk}. For fixed central charge, the space of such forms is finitely generated \cite{9781468491647}, and moreover they are protected in moduli space. It is thus possible to systematically classify elliptic genera compatible with growth of the spectrum required by holography, as was done in \cite{Benjamin:2015vkc,Belin:2016knb,Belin:2018oza,Belin:2019jqz,Belin:2019rba}. One striking outcome was the condition that the seed theory $\cal C$ of the symmetric product orbifold needs to have central charge $c\leq6$ \cite{Belin:2019jqz}. 

Building on this result, \cite{Belin:2020nmp} proposed an infinite new family of holographic CFTs, namely symmetric product orbifolds of ${\cal N}=2$ minimal models. So far, there were two pieces of evidence for their holographic nature. First, they all have at least one exactly marginal single-trace operator in the twisted sector.  Second, using \cite{Belin:2019jqz,Belin:2019rba,Keller:2020rwi}, one can show that the growth of degeneracies in the elliptic genus of the orbifold theory is compatible with a supergravity description in AdS$_3$. Even though these degeneracies are computed at the orbifold point, they are invariant on the conformal manifold and hence probe the putative strongly-coupled supergravity point. 

One of the goals of the present paper is to provide further evidence for the existence of a holographic CFT on the moduli space of these symmetric product orbifolds.
To explain this, let us
discuss how such theories could fail to be actually holographic. At the orbifold point, the large gap condition is violated, as there are infinite towers of higher spin currents. For the large gap condition to be satisfied elsewhere in the moduli space, it is thus necessary that these currents are lifted and acquire an anomalous dimension as one moves away from the orbifold point by turning on $\lambda$. If this does not happen for all higher spin currents, the large gap condition cannot be satisfied. 

The general lore is that all operators which are not protected should get lifted under the deformations, but this is of course only an expectation and is far from being a theorem. In this paper, we will study deformations of symmetric product orbifolds perturbatively in $\lambda$ and apply conformal perturbation theory to show that the non-protected currents do indeed acquire an anomalous dimension already at order $\lambda^2$.

Let us note that even though our results provide supporting evidence, it is possible that the anomalous dimensions saturate at an $\mathcal{O}(1)$ value as we try to take $\lambda\to\infty$. Whether or not the anomalous dimensions saturate is difficult to analyze as it entails understanding the theories non-perturbatively, for which very few tools are available and which is out of our scope. The analysis here is aimed to explore if there are any surprises within the perturbative regime of $\lambda$.

\subsection{Summary of results}

In this paper we present a general analysis regarding the effects of deforming symmetric product orbifolds of two-dimensional $\cN=2$ CFTs by an exactly marginal operator. We consider higher spin currents constructed from the $N$ copies of the $\cN=2$ algebra present at the symmetric orbifold point, and investigate how they are lifted under deformations; that is, we compute the anomalous dimension they acquire. 

For practical reasons we only consider currents of spin 2, 5/2 and 3. Our methods however would work just as well for higher spin currents, and we provide a general procedure to construct them in Sec.\,\ref{sec:symn}, where we also review general aspects of symmetric orbifolds. When evaluating the anomalous dimensions, we focus exclusively on currents of ${\rm Sym}^N({\cal C})$ that are constructed out of the superconformal algebra of the seed theory. Incorporating additional currents that are specific to enhanced symmetries that might appear in  ${\cal C}$ is doable, but is not considered in the analysis presented here. For the infinite family of the A-series $\cN=2$ minimal models there are no such additional currents, and thus for these seed CFTs our analysis is exhaustive.

In order to evaluate the anomalous dimensions we use conformal perturbation theory, which we review in Sec.\,\ref{sec:PertCFT}. More precisely, we compute the anomalous dimensions at second order in perturbation theory. Conceptually, this is a straightforward task; technically, it turns out to be a quite difficult computation. The main issue is that the marginal operators used to deform the theory live in the twisted sector, so we need to evaluate twisted correlation functions. To do this, as is standard, we map the correlators to a covering surface, and then use Ward identities to evaluate the correlators on the cover. Although this is in principle also straightforward, in practice the computation quickly becomes cumbersome due to the number of currents involved. The technical aspects of these computations are described in Sec.\,\ref{sec:coverbasemap}.

Even though the anomalous dimensions for higher spin currents are primarily computed using Ward identities, we discover that the lifting mechanism is surprisingly non-universal. First, the twist of the marginal operator depends on the central charge of the seed theory. We give a complete list of such possible operators for any seed theory ${\cal C}$ with $1\leq c\leq 6$.\footnote{In the range $1\leq c< 3$ the list is exhaustive for unitary and discrete CFTs as these correspond to ${\cal N}=2$ minimal models. For $3\leq c\leq6$ our results predict the primary operator that must be present in the seed theory for the twisted moduli to exist.} Second, we find that although the moduli are by definition primary fields on the base, their images on the cover are not necessarily primaries. This leads to another non-universal aspect of the deformation which explicitly affects the magnitude of the anomalous dimensions.

Despite the aforementioned non-universal aspects of the computation, the overall outcome is completely universal: we find that every higher spin current that is expected to be lifted, does get lifted. These results are presented in Sec.\,\ref{sec:lifths}. The computations are valid for finite values of $N$, and we take the large-$N$ limit to make comparisons with prior literature --- which only exists for $c=6$ when the seed is ${\mathbb{T}^4}$.

\section{Conformal perturbation theory}\label{sec:PertCFT}

We start by reviewing the salient features of conformal perturbation theory for two-dimensional conformal field theories.

\subsection{General setup}\label{sec:generalCFT}

Conformal perturbation theory can be used to explore the moduli space of a CFT containing an exactly marginal operator $\Phi$, \ie an operator with scaling dimension $(h,\bar h)=(1,1)$ whose weight remains unchanged to all orders in perturbation theory. To deform the CFT with this marginal operator we add the following term to the action
\be
\Delta S\coloneqq\lambda\int d^2z\ \Phi(z,\bar z)\,. 
\ee
Because $\Phi$ is exactly marginal, the deformed theory is again a CFT. Normalized correlation functions in the deformed theory can be computed order by order by expanding the exponential\footnote{In the following we suppress the anti-holomorphic coordinates when possible.}
\begin{equation}\label{eq:exp}
    \begin{aligned}
\left\langle O_a(z) O_b(z')\right\rangle_{\lambda}\coloneqq&\,\frac{\big\langle O_a(z) O_b(z') e^{\lambda\int d^2w \ \Phi(w)}\big\rangle}{\left\langle e^{\lambda \int d^2w\ \Phi(w)}\right\rangle}\\
=&\left\langle O_a(z) O_b(z')\right\rangle+\lambda\int d^2w \ \left\langle  O_a(z) O_b(z')\Phi(w)\right\rangle_{\textrm{connected}}\\
&\hspace{20pt}+\frac{\lambda^2}{2}\int d^2w \ d^2w' \ \left\langle  O_a(z) O_b(z')\Phi(w)\Phi(w')\right\rangle_{\text{connected}} +\cO(\lambda^3)\,.
    \end{aligned}
\end{equation}
Here $O_{a,b}$ are primaries of the undeformed theory of weight $h_{a,b}$ that we assume to be orthogonal to one another. 

A typical question we ask is, how does the conformal weight $h_a$ of an operator change under this deformation? To answer this question, we note that the deformed theory is again a CFT, so the normalized two-point functions are completely fixed by symmetry to be of the form
\begin{align}
\left\langle \mathbb{O}_a(z) \mathbb{O}_a(z')\right\rangle_{\lambda}&=\frac{1}{(z-z')^{2(h_a+\mu_{a}(\lambda))}(\bar z-\bar z')^{2(\bar h_a+\bar\mu_{a}(\lambda))}}\,. \label{eq:deformation}
\end{align}
Here $(h_a,\bar h_a)$ is the scaling dimension of the undeformed operator, and $(\mu_{a}(\lambda),\bar \mu_{a}(\lambda))$ is its correction induced by the deformation. Note that the operators $\mathbb{O}_a$ are not necessarily the same as the operators $O_a$ as they have been rotated into a basis where the two-point functions are diagonal. Locality then forces $\mu_{a}=\bar \mu_{a}$ so that, in particular, the spin remains unchanged by the deformation.  For the purpose of this work, we will only be interested in lifting holomorphic operators $(\bar h_a=0)$. 

To relate (\ref{eq:deformation}) to (\ref{eq:exp}),
we note that evaluating the integrals in the latter will lead to an expression of the form
\be\label{eq:ad}
\langle O_a(z) O_b(z')\rangle_{\lambda}=\frac{1}{(z-z')^{2h_a}}\big(\delta_{ab}-2\gamma_{ab}\log\abs{z-z'}^2+ \ldots \big)\,.
\ee
In this expression, $\gamma_{ab}$ denotes a mixing matrix that captures the effects of the deformation: its eigenvalues give the anomalous dimensions $\mu_a$, and its eigenbasis describes the operator mixing that relates $\mathbb{O}_a$ to $O_a$. Therefore, one can determine the anomalous dimensions of operators by computing the correction terms in (\ref{eq:exp}) and extracting the coefficient of the logarithm. 

We will be interested in the case when the CFT that we deform is a large-$N$ symmetric product orbifold of an $\cN=2$ minimal model. These theories have many exactly marginal operators \cite{Belin:2020nmp}. We consider marginal operators coming from a single-trace $1/2$-BPS operator in the  twisted sector. More concretely, the marginal operators will be $G$-descendants of a single-trace (anti-)chiral primary in the twisted sector (see App.\,\ref{app:conventions} for our conventions and the definition of the $\mathcal{N}=2$ algebra). The fact that the deformation operator is $1/2$-BPS ensures that the deformed theory preserves $\cN=2$ supersymmetry. Moreover, because of this, the marginal operator itself is protected from getting an anomalous dimension, guaranteeing that it is marginal to all orders in perturbation theory \cite{Dixon:1987bg,Gukov:2004ym}.

The fact that the marginal operator is in the twisted sector on the other hand causes the orbifold structure to be broken as we turn on $\lambda$. This is actually a promising feature: symmetric orbifolds universally exhibit a Hagedorn growth in their density of states \cite{Keller:2011xi}, and hence cannot be dual to supergravity theories in AdS$_3$ (in fact, they are dual to tensionless string theory --- see \cite{Eberhardt:2018ouy} for the symmetric orbifold of $\mathbb T^4$). Turning on the deformation may push the bulk dual into the supergravity regime by making the string states heavy. In other words, the existence of such marginal operators make the existence of a supergravity point in the moduli space possible. 

The reason to consider single-trace moduli is more subtle. In the setting of symmetric orbifolds, the deformation can be seen as turning on a gauge coupling with an 't Hooft like expansion at large $N$. As we will review in Sec.\,\ref{sec:coverbasemap}, it turns out that to have a good planar limit as $N\rightarrow\infty$ we must dress the $\cO(1)$ coupling constant $\lambda$ with an appropriate power of $N$
\be
\lambda\mapsto\lambda N^\frac{1}{2}\,.
\ee
By exploiting this fact, it can be shown that to leading order in $N$ only single-trace moduli can cause significant changes in the spectrum (see App.\,D of \cite{Belin:2020nmp}).

\subsection{Higher spin currents}\label{sec:lc}

In this work we will be interested in the lifting of the higher spin currents in the untwisted sector of the orbifold theories. This will provide a number of simplifications that we explain here. Let us first provide the necessary notation. 

Naturally, since we consider $\cN=2$ superconformal theories, the operators (and thus the currents) will organize themselves into multiplets. We will use ${\cal W}_{s}$ to denote the superprimary field as defined in App.\,\ref{app:superprimary}; its components will be denoted by
\begin{equation}\label{eq:superHS}
    \mathcal{W}_{s}\coloneqq \left\{W_s, \mathscr{W}_{s+1/2}^+, \mathscr{W}_{s+1/2}^-, \mathtt{w}_{s+1}\right\}\,.
\end{equation}
We will report explicitly on the anomalous dimension of $W_s$. The corresponding corrections for the other entries of the multiplet are determined in a straight forward manner by supersymmetry. 
If there is more than one current at fixed spin $s$, we will use the notation $\mathcal{W}_{s,a}=\big\{W_{s,a},\ldots \big\}$. For very low values of $s$, and under the assumption that the seed theory has no more than an ${\cal N}=2$ superconformal symmetry, there is only one superprimary current at fixed $s$.  But as the spin increases, there will be several multiplets for fixed $s$, and these should be distinguished appropriately. 

There are additional operators, which mimic spin-$s$ currents, but are merely descendants of the global superconformal algebra. For the simplest examples in Sec.\,\ref{sec:lifths} we will highlight them for pedagogical reasons, and they will be denoted  by $Y_{s}(z)$. 

From (\ref{eq:exp})  we see that at linear order in $\lambda$ the lifting of the currents will be determined by a three-point function of the form
\be
\left\langle W_{s,a}(z)W_{s,b}(z')\Phi(w)\right\rangle\, ,
\ee
with $W$ a current in the untwisted sector, and $\Phi$ a twisted sector modulus. This three-point function must vanish for two reasons. First of all, the twist selection rule, \ie the criterion that the elements $g\in S_N$ determining the twists of the twisted sector operators in a correlation function must multiply to unity to yield a non-zero result, ensures that any correlation function with only one twisted sector insertion vanishes. The second argument is more general and applies to any marginal operator and any CFT: for the lifting of holomorphic currents, the first order correction is given by a three-point function involving two currents and a scalar --- the modulus --- which has $\bar h=1$. Such correlation functions always vanish in CFTs. This is closely tied to unitarity: if currents could acquire a first-order anomalous dimension, one could give them a negative anti-holomorphic weight $\bar{h}$ by appropriately choosing the sign of $\lambda$, which violates unitarity. 

The first non-vanishing contribution to the anomalous dimension of currents thus comes at order $\lambda^2$, and can be found by computing a four-point function of the form
\be\label{basecorrelator}
\left\langle W_{s,a}(z)W_{s,b}(z')\Phi(w)\Phi(w')\right\rangle\, .
\ee
Computing these four-point functions and extracting the anomalous dimensions will be the focus of the rest of this work.

We note that in general we will have to perform degenerate perturbation theory, taking into account operator mixing. To reduce the size of the lifting matrices, we can use the fact that the spin and the $U(1)$ charge of the operators are preserved, meaning that any matrix elements between operators of different spin or charge vanishes. We can also focus on primary fields only, since those determine the lifting of their descendants.
For instance,
at the lowest spin, $s=2$, there is only one current, so that the anomalous dimension is determined by a single number. For the next value of the spin, $s=3$, the situation is already a bit more complicated. In that case, we will need to compute the eigenvalues of a two-by-two matrix as the mixing matrix $\gamma_{ab}$ is non-trivial. 

Let us conclude this subsection by describing some consistency requirements and tricks to simplify the computation of anomalous dimensions of holomorphic currents. First of all,
whenever one computes the anomalous dimensions of holomorphic operators, the eigenvalues of the lifting matrix must be non-negative. This follows from the fact that the spin $s=h-\bar h$ is conserved along the deformation. Hence, a negative eigenvalue in the lifting matrix would imply a negative anti-holomorphic weight $\bar h$ once $\lambda$ is turned on.

In addition to that, a strategy that we will employ many times, and that simplifies our calculations considerably, is based on the following. Since the deformation operator is exactly marginal and preserves supersymmetry, the $\cN=2$ algebra of the symmetric orbifold will not be broken under deformation. Therefore, operators corresponding to the diagonal $\cN=2$ algebras do not get lifted at any order in perturbation theory; the $Y_s$ defined below will be examples of such operators protected by superconformal symmetry. We can use this fact extensively because the higher spin currents are (partly) built out of precisely such operators. Thus, the lifting of the currents must be originating purely from the terms that come from the algebra of the seed, but are not protected by the algebra of the orbifold.

\subsection{Integrals and regularization}\label{sec:i&r}

In this subsection we will discuss how to extract the anomalous dimension from a four-point function between two currents and two moduli. We will mostly follow \cite{Gaberdiel:2015uca,Keller:2019suk,Keller:2019yrr}. Our starting point is the first non-zero term in the expansion (\ref{eq:exp}), which we showed to be the term at order $\lambda^2$ given by
\be
\frac{\lambda^2N}{2}\int d^2w \ d^2w' \ \left\langle  W_{s,a}(z) \Phi(w)\Phi(w')W_{s,b}(z')\right\rangle\label{eq:lsq}\, .
\ee
This integral is UV-divergent. However, we can extract the anomalous dimension by splitting the integral into a UV-finite part that contains the anomalous dimension and a divergent part. This can be done as follows. We simplify the integral by using global conformal symmetry to write the four-point function as a function of a cross-ratio. To that end, we use the following M\"{o}bius transformation that maps a complex  coordinate \nolinebreak$u$ \nolinebreak to
\eq{
u \mapsto  f(u)\coloneqq \frac{(u - z')(w - z)}{(u - z)(w - z') - (u - w)( z - z')\delta}\,,\label{eq:ct}
}
that sends the insertions located at $z'$, $w$, and $z$ to 0, 1, and $1/\delta$ respectively, and sends $w'$ to a cross ratio that will be specified shortly. Here taking $\delta\to0$ should be understood as a regulator that will be used to define a normalized bra state. Implementing this transformation on the four-point function and keeping only the leading order in $\delta$ in \eqref{eq:lsq} 
results in
\begin{align}
\begin{split}
&\big\langle  W_{s,a}(z) \Phi(w)\Phi(w')W_{s,b}(z')\big\rangle\\
&\hspace{60pt}=\frac{1}{\delta^{2s}(z-z')^{2s}}\left\vert\frac{(z-z')^2}{(w'-z)^2(w-z')^2}\right\vert^2\left\langle W_{s,a}(\delta^{-1})\Phi(1)\Phi(x)W_{s,b}(0)\right\rangle\, ,\label{eq:ct2}  
\end{split}
\end{align}
with $x$ the cross-ratio given by
\be
x\coloneqq \frac{(w'-z')(w-z)}{(w'-z)(w-z')}\, .
\ee
Subsequently, we perform a change of variables so that the integration is over the coordinates $w$ and $x$. The measure of integration then changes as follows
\begin{align}
\begin{split}
d^2w\ d^2w'&=d^2w\ d^2x\left\vert\frac{\partial(w,w')}{\partial(w,x)}\right\vert^2 =d^2w\ d^2x\left\vert\frac{(w'-z)^2(w-z')}{(w-z)(z-z')}\right\vert^2\, .\label{eq:jac}
\end{split}
\end{align}
By combining (\ref{eq:ct2}) and (\ref{eq:jac}) we are left with the following integrals to compute
\begin{align}
\int d^2w\ d^2x \ \frac{1}{\delta^{2s}(z-z')^{2s}}\left\vert\frac{z-z'}{(z-w)(w-z')}\right\vert^2 \left\langle W_{s,a}(\delta^{-1})\Phi(1)\Phi(x)W_{s,b}(0)\right\rangle \label{eq:mobtransf}\,.
\end{align}
We see that the regulator $\delta$ can be absorbed into the definition of a properly normalized bra state $\bra{W_{s,a}}$. 

At first glance we could now perform the integration over $w$ without knowing the four-point function. 
Note though that the cross-ratio $x$ depends on $w$. Therefore, in principle one has to be mindful of the $w$-dependence of $x$ when evaluating the $w$ integral in (\ref{eq:mobtransf}). Nevertheless, it turns out that as long as
\eq{
\bar h_W\neq \bar h_\Phi\, ,
}
the logarithmic term that we are after can be found by evaluating the $w$ integral as if there was no $w$-dependence in $x$.\footnote{See \cite{Keller:2019suk,Keller:2019yrr} for a proof of this result.} Hence, since $\bar h_W=0$ and $\bar h_\Phi=1$, we can perform the $w$ integral by cutting out $\epsilon$-discs around $z$ and $z'$ to arrive at
\be
\frac{2\pi}{(z-z')^{2s}}\log{\bigg(\frac{\abs{z-z'}^2}{\epsilon^2}\bigg)}\int d^2 x\ \bra{W_{s,a}}\Phi(1)\Phi(x)\ket{W_{s,b}}\, .
\ee
The anomalous dimension can thus be found by computing
\be\label{eq:holintegral}
\gamma_{ab}=-\frac{\pi\lambda^2N}{2}\int d^2 x\ \bra{W_{s,a}}\Phi(1)\Phi(x)\ket{W_{s,b}}\, .
\ee
The final remaining obstacle is evaluating the $x$ integral. Once again, this integral is divergent. We can regulate it by cutting out $\epsilon$-discs around the points $0$, $1$, and $\infty$. Since the final result should be scheme independent, the anomalous dimension will be given by the constant term in the $\epsilon$ expansion. 

We can now use the fact that the moduli $\Phi$ are $G$-descendants of 1/2-BPS states to convert the area integral in (\ref{eq:holintegral}) into a contour integral: we use the Ward identities for the  $\bar G_{-1/2}$ to convert them into anti-holomorphic derivatives. Stokes' theorem then gives contour integrals around $0$, $1$, and $\infty$. It turns out that the integral around $1$ does not contribute to the anomalous dimension. Schematically this can be seen from the fact that this term comes from contracting the two moduli together. Once all the dust settles all that remains of the integration is a contour integral around $0$ and $\infty$
\be\label{eq:finalgamma}
\gamma_{ab}=-\frac{i\pi\lambda^2N}{4}\left(\oint_0 dx\ -\oint_\infty dx \ \right)\bra{W_{s,a}}\hat\Phi(1)\hat\Phi(x)\ket{W_{s,b}}\, ,
\ee
where we used $dx^2=\frac{i}{2}dxd\bar x\,$, and $\hat\Phi$ denotes the operator $\Phi$ without $\bar G_{-1/2}$ descendants. More precisely, since $\Phi$ has to be Hermitian, we can write it as the sum of an operator and its Hermitian conjugate; $\hat \Phi$ is then decomposed into its chiral ($\varphi_c$) and anti-chiral ($\varphi_a$) components via
\be
\ket{\hat\Phi}=\frac{1}{\sqrt{2}}\big(G^{-}_{-1/2}\ket{\varphi_c}+G^{+}_{-1/2}\ket{\varphi_{a}}\big)\, .\label{eq:chiralmariginal}
\ee
This will lead to four possible contributions to the anomalous dimensions. However, charge conservation ensures that just the terms with one chiral and one anti-chiral insertion can contribute. It turns out that at the level of correlation functions, the two non-zero contributions do not have to agree. They do however give equal contributions to the anomalous dimensions. Computing these correlation functions will be the focus of the following sections.

\section{Structural aspects of symmetric product orbifolds}\label{sec:symn}

In this section we highlight properties of symmetric product orbifolds that will be used in subsequent sections. We will review the effects of the symmetric group on the spectrum and correlation functions, and introduce the terminology that we will use to describe the states in the theory. Along the way we will review how these features lead to a well-defined large-$N$ limit in Sec.\,\ref{sec:largeN}. Thereafter, we will discuss the appearance of spin-$s$ currents in symmetric product orbifolds and their behaviour at large $N$. 

Some basic properties of symmetric product orbifolds are the following. Symmetric product orbifolds are obtained by taking $N$ copies of a seed CFT $\mathcal{C}$ and orbifolding by the $S_N$ permutation symmetry between the copies
\be
{\rm Sym}^N({\cal C})\coloneqq \frac{ \mathcal{C}^{\otimes N}}{S_N} \,.
\ee
We will refer to ${\cal C}$ as the \textit{seed} theory, and denote its central charge with $c$. The central charge of Sym$^N({\cal C})$ is therefore $Nc$.

The orbifolding produces two effects: First, it projects the Hilbert space of the product theory down to $S_N$ invariant states. We will refer to this sector as the \textit{untwisted} sector. Second, modular invariance of the theory requires the addition of new states, known as \textit{twisted} sectors. In these sectors, the boundary conditions of the fields are twisted by the action of $S_N$. Twisted sectors are labelled by conjugacy classes of the group, which for $S_N$ are in one to one correspondence with the partitions of $N$ (or Young diagrams). The twisted sectors that will be relevant for this paper are simple, and correspond to a single non-trivial row of length $n$. We will refer to this sector as the \textit{twist}-$n$ sector.

Let us first discuss operators in the untwisted sector and their correlation functions. We can think of operators in the untwisted sector as given by orbits under the action of $S_N$ that permutes the $N$ factors. That is, we can start with a state in the $N$-fold tensor product theory of the form 
\be\label{prestate}
O^{(1)}_{a_1} \, \otimes  \cdots \otimes\, O^{(l)}_{a_l}\,\otimes\, \underbrace{\mathbb{1}\, \otimes \cdots \otimes \,\mathbb{1}}_{N-l}\ .
\ee
Here the upper label $(i)$ denotes the different tensor copies while the lower label $a_i$ denotes the operators in the seed theory. An operator in the untwisted sector of the symmetric orbifold is then a symmetrized version of this,
\begin{align}\label{prestate1}
\begin{split}
O &= {\rm Sym}(O^{(1)}_{a_1} \, \otimes  \cdots \otimes\, O^{(l)}_{a_l}\,\otimes\, \underbrace{\mathbb{1}\, \otimes \cdots \otimes \,\mathbb{1}}_{N-l})\\
&\hspace{160pt}= \sum_{g\in S_N} g\cdot(O^{(1)}_{a_1} \, \otimes  \cdots \otimes\, O^{(l)}_{a_l}\,\otimes\, \underbrace{\mathbb{1}\, \otimes \cdots \otimes \,\mathbb{1}}_{N-l} )\ .
\end{split}
\end{align}
Since we can use (\ref{prestate}) to construct such a state for any $N\geq l$, this allows us to construct a large-$N$ limit. 

Computing $n$-point correlation functions of such untwisted operators is then straightforward: the result is simply the product of the correlators of the individual factors, summed over $l$ copies of $S_N$. The only slightly delicate issue is to keep track of this sum and the corresponding normalization --- see App.\,\ref{app:largeN} for some remarks on this.

As mentioned above, a state in the twisted sector has a simple geometrical interpretation: it corresponds to an operator with a modified boundary condition of the form
\be\label{eq:twistact}
O(e^{2\pi i}z)^{(j)}= O(z)^{g(j)}~, \qquad  g\in S_N~.
\ee
Note that this boundary condition is only determined by the conjugacy classes of $S_N$. In this context a \textit{twist field}, $\sigma_g(z)$, is defined as an operator that captures the non-trivial monodromy induced by (\ref{eq:twistact}). What we denote as the twist-$n$ sector, for which we have a twist field $\sigma_n(z)$, are the cases where the permutation is a single cycle of length  $n$. In a given twist sector, $\sigma_n(z)$ is the lowest dimension operator of that sector, and other states in the sector come from excited states of the seed theory, uplifted to the twist sector.  

Computing correlation functions of twisted fields is more complicated than correlation functions of untwisted fields: twist fields, when inserted in a correlation function, modify the periodicity of operators around the point where they are inserted.
The basic idea to overcome this complication is to map the correlator to an $n$-sheeted cover, where each sheet corresponds to a tensor factor of the orbifold CFT \cite{Lunin:2000yv,Lunin:2001pw}. This undoes the effect of the twist by making fields again single-valued on the covering space. If there are more than two twist operators in the correlation function, the covering space is typically a higher genus surface. Fortunately, this will not be the case for the correlators appearing in the rest of this paper. Evaluating correlation functions involving twist fields will be an important aspect of our computation and it will be discussed at length in Sec.\,\ref{sec:coverbasemap}.

In the presence of fermions, the definition of the twist field needs to be expanded to incorporate them. The basic reason is the non-trivial interplay between (\ref{eq:twistact}) and the NS and R boundary conditions as we map objects from the base to the cover. This is carefully explained in \cite{Lunin:2001pw}, and will also be addressed in  Sec.\,\ref{sec:coverbasemap}.

\subsection{Large-\texorpdfstring{$N$}{N} limit}\label{sec:largeN}

The behaviour of symmetric product orbifolds at large $N$ is reminiscent of standard gauge theories with a 't Hooft limit. Here, we summarize the features of this limit relevant for this work; see \cite{Pakman:2009zz,Belin:2015hwa} for more details. The main property that will be important in this paper is large-$N$ factorization, which impacts our choice of the deformation of Sym$^N({\cal C})$ via $\Phi$: the deformation should persist in the limit $N\to\infty$.  

To characterize the limit and its inner-workings, it is useful to decompose operators in symmetric orbifolds into single and multi-trace operators. The outcome will be very similar to large-$N$ properties of gauge theories.  We will mainly concentrate on the untwisted sector here, since explicit expressions are simpler. The structure and combinatorics for the twisted sector are similar.

\paragraph{Single-trace operators.} We say that an operator of the form (\ref{prestate1}) has trace length $l$. In particular, a single-trace operator has $l=1$ and can be written as
\be
O = (N!(N-1)!)^{-1/2}\sum_{g\in S_N} g\cdot  O^{(1)} = N^{-1/2}\sum_{i=1}^N O^{(i)}~,
\ee
where, by a slight abuse of notation, $O^{(i)}$ denotes the operator with $O$ in the $i$-th copy and the vacuum everywhere else. We have given the explicit normalization factor since it will play a crucial role in the factorization described below. In particular, this normalization assures that $\braket{O O}\sim {\cal O}(N^0)$.
In the twisted sector, a single-trace operator has a single cycle of length more than 1, and all other factors are in the vacuum.

Large-$N$ factorization is the statement that correlation functions of single-trace operators factorize in the large-$N$ limit. For instance, given four single-trace operators $O_{1,\cdots,4}$ we have
\be\label{eq:4ptN}
\braket{O_1 O_2 O_3 O_4}=\braket{O_1 O_2}\braket{O_3 O_4} + \braket{O_1 O_3}\braket{O_2 O_4}+\braket{O_1 O_4}\braket{O_2 O_3} +\mathcal{O}(1/N) \,.
\ee
That is, the leading contribution comes from disconnected two-point functions. This relation also holds for both the twisted and untwisted single-trace states of Sym$^N({\cal C})$. The large-$N$ decomposition has important consequences for the computations we will perform in Sec.\,\ref{sec:coverbasemap}.

\paragraph{Multi-trace operators.} 
Let us briefly describe multi-trace operators and their large-$N$ limit.
Multi-trace operators are built from operators (\ref{prestate}) with $l>1$, such as 
\be
O \sim \sum_{g\in S_N} g\cdot(O^{(1)}_{a_1} \, \otimes  \cdots \otimes\, O^{(l)}_{a_l}\,\otimes\, \underbrace{\mathbb{1}\, \otimes \cdots \otimes \,\mathbb{1}}_{N-l} )\ .
\ee
In the twisted sector, a multi-trace operator corresponds to multiple non-trivial cycles, or one non-trivial cycle together with non-vacuum operators in some other factors. 

It can happen that it is more convenient to work with operators which have both single and multi-trace parts. Certain operators need to have both single and multi-trace components to be primary operators, for example the higher spin currents that we will discuss below. Such operators are of the form
\be
O \sim \sum_{g\in S_N} g \cdot (O_{st}+O_{mt})\ .
\ee
To leading order in large $N$, the normalization of $O$ is then given by the normalization of the single-trace part $O_{st}$. In fact, the leading order term of the correlation functions of such fields is also given just by the single-trace contribution. For a more detailed discussion of this see App.\,\ref{app:largeN}.

Another example of fields with different trace length components are products of single-trace operators such as
\be
OO \sim (\sum_{g\in S_N} g \cdot O^{(1)})(\sum_{g\in S_N} g \cdot O^{(1)}) \sim \sum_{g\in S_N} g \cdot(O^{(1)}O^{(1)} + (N-1) O^{(1)}O^{(2)})
\ee
Here there is an explicit $N$ dependent factor between the two terms, so that the single and multi-trace part contribute equally to the normalization and the correlation functions.

\subsection{Construction of currents in \texorpdfstring{Sym$^N({\cal C})$}{SymN(C)}}\label{sec:currents}

In this subsection we will discuss the basic ingredients and structures that appear in the construction of higher spin currents in Sym$^N({\cal C})$. We will start with a fairly general analysis that will be independent of the seed theory ${\cal C}$. This analysis relies on the (super-)conformal symmetries of the seed, but not on its primary operator content for the moment. One of our working assumptions here is that the seed theory has at most ${\cal N}=2$ superconformal symmetry.    

Starting with the universal sector, the currents of the seed theory are given by a $\cN=2$ superconformal algebra. The stress tensor, supercurrents, and the $U(1)$ current of Sym$^N({\cal C})$ are therefore described by fields in the untwisted sector; these can be written as
\begin{equation}
\begin{aligned}
T(z) &= {\rm Sym}(T^{(1)}(z) \otimes \mathbb{1} \,\otimes \mathbb{1}\, \dots ) = \sum_{i=1}^N T^{(i)}(z)\,, \label{stresstensor} \\
J(z) &=  {\rm Sym}(J^{(1)}(z) \otimes \mathbb{1} \,\otimes \mathbb{1}\, \dots ) =  \sum_{i=1}^N J^{(i)}(z)\,, \\
G^\pm(z) &= {\rm Sym}(G^{\pm(1)}(z) \otimes \mathbb{1} \,\otimes \mathbb{1}\, \dots ) =  \sum_{i=1}^N G^{\pm(i)}(z)\, ,
\end{aligned}    
\end{equation}
where $T^{(i)}(z)$, $J^{(i)}(z)$, and $G^{\pm(i)}$ are the stress tensor, $U(1)$ current, and supercurrents in the $i^{th}$ copy of the CFT.\footnote{Note that we have \textit{not} normalized these fields such that the coefficient of their two-point functions is one, as is standard for currents.}

Let us start with the design of a spin-4 current in Sym$^N({\cal C})$. This example will serve as a representative to highlight the basic structures.  In any CFT$_2$  there is a spin-four field constructed from the stress tensor that is given by 
\begin{equation}
Y_{4}(z) = (T T)(z)- \frac{3}{10} \p^2 T(z)\,, \label{O4}    
\end{equation}
where $(AB)(z)$ denotes the normal-ordered product of two operators $A(z)$ and $B(z)$. The field $Y_4(z)$ is not a primary, however, as can be seen from the OPE
\begin{equation}
T(z) Y_4(w) =  \frac{(\tfrac{22}{5} + N c)T(w)}{(z-w)^4} + \frac{4  Y_4(w)}{(z - w)^2} + \frac{\p Y_4(w)}{z-w}\,. \label{opeTO4}
\end{equation}
For a symmetric orbifold, $Y_4(z)$ can be turned into a primary field by noting that there is another spin-4 field constructed from the stress tensor. When $i \ne j\,$, we have
\eq{
T^{(i)}(z) (T^{(i)} T^{(j)}) (w) = \frac{\tfrac{c}{2} T^{(j)}(w)}{(z-w)^4} + \frac{2 (T^{(i)} T^{(j)})(w)}{(z-w)^2} + \frac{(\p T^{(i)} T^{(j)})(w)}{z - w}\,, \qquad i \ne j\,. \label{newope}
}
Using \eqref{newope} it is not difficult to show that any symmetric orbifold CFT features a primary of weight four that is given by
\begin{align}
  W_4(z) &= (T T)(z) - \frac{3}{10} \p^2 T(z) - \frac{\tfrac{22}{5c}+ N}{N- 1}\sum_{i\ne j} (T^{(i)} T^{(j)})(z), \label{W4} \\
& = \sum_{i=1}^N  \Big[ (T^{(i)} T^{(i)})(z) - \frac{3}{10} \p^2 T^{(i)}(z) \Big] -  \frac{\tfrac{22}{5c} + 1}{N- 1}\sum_{i\ne j} (T^{(i)} T^{(j)})(z)\,.
 \label{W42}     
\end{align}
We are casting $W_4$ in two equivalent ways, \eqref{W4} and \eqref{W42}, to make two features manifest: In \eqref{W4}, the first two terms are protected since they are composed of stress tensors of the whole CFT, and therefore do not acquire an anomalous dimension under the deformation; we will make this explicit in the computations of Sec.\,\ref{sec:lifths}. Only the last term acquires an anomalous dimension. In \eqref{W42}, we split $W_4(z)$ into single and double-trace terms. This is useful for understanding how the currents behave in the large-$N$ limit, and how to organize subsequent computations efficiently in this limit. That is,  if we only cared about $N\to\infty$, then only the first two single-trace terms are important and we can neglect the double-trace term. 

In subsequent sections we will just present the currents for the specific values of spin we have analysed following the principle we have presented here for $W_4$, that is constructing primary operators built from symmetrized currents living in the different copies of the seed CFT. The only nuance is that we have to include the $U(1)$ current and fermionic symmetries of the $\cN=2$ superconformal algebra in the construction of the currents. More concretely, the procedure is:
\begin{enumerate}
    \item At a fixed spin $s$, we list all possible combinations of $T^{(i)}$, $J^{(i)}$, and $G^{\pm (i)}$, including derivatives, such that the total weight adds up to $s$. 
    \item We classify which of those combinations are primaries with respect to the ${\cal N}=2$ algebra.\footnote{This is a more stringent condition than requiring the field to be just a Virasoro primary as we did for $W_4$.} 
    \item  We organize the currents into multiplets, according to the ${\cal N}=2$ algebra, see  \eqref{eq:superHS}. 
    \item Finally, we normalize the operators such that their two-point functions are unity in accordance with \eqref{eq:ad}.\footnote{Note that in \eqref{W4} and \eqref{W42} we have not normalized $W_4$. } 
\end{enumerate}

In this article we only consider currents built from $\cN=2$ superconformal currents. However, in principle the seed theory ${\cal C}$ can have additional currents. This happens, for example, when $\mathcal{C}$ is a free theory; the most studied example being the sigma model on $\mathbb T^4$. Even though we do not consider such examples, the overall procedure we have presented here remains the same; the only difference is to incorporate for step 1 those additional currents to the list of possible combinations, and possibly arrange the currents into primaries and multiplets with respect to a larger symmetry algebra.

\section{Deformations of symmetric product orbifolds}\label{sec:coverbasemap}

Let us now discuss how to deform symmetric product orbifold CFTs via a marginal deformation.  To do this, we first need to choose an exactly marginal operator $\Phi$. Let us recap the arguments of Sec.\,\ref{sec:PertCFT} that constrain our choices, and place them in the context of the structural aspects of symmetric product orbifolds discussed in Sec.\,\ref{sec:symn}. 

In order to produce the desired effect, we want the marginal operator $\Phi$ to be:
\begin{description}[leftmargin=0cm]
    \item[1/2-BPS:] On the one hand this guarantees that the deformed theory preserves $\cN=2$ supersymmetry. On the other hand it guarantees that $\Phi$ is protected from getting an anomalous dimension, and therefore remains marginal to all orders in perturbation theory.
    \item[Twisted:] This breaks the orbifold structure of the theory. That is, the deformed theory has interactions among the copies of Sym$^N(\cal{C})$, which could allow for strong coupling effects controlled by $\lambda$.
    \item[Single-trace:] We are most interested in lifting the currents $W_{s}$ in Sec.\,\ref{sec:currents}, which to leading order in $N$ are single-trace.\footnote{In general, we want most states to lift in order to meet the sparseness conditions of a dual low-energy effective field theory. We are focusing on currents because they exist for any Sym$^N(\cal{C})$ theory, and if they persist it is problematic for the local properties of the dual field theory.}  From the discussion in Sec.~\ref{sec:largeN}, if $\Phi$ is  single-trace then the connected pieces scale like $1/N$. Consequently, in order to produce anomalous dimensions that are $\mathcal{O}(1)$ in the large-$N$ limit, we need to deform the CFT by\footnote{We are assuming that the deformation operator is normalized to have unit two-point function.}
\be
S\to S + \lambda \sqrt{N} \int d^2z \ \Phi(z,\zb)\ \,.
\ee
This is the scaling that preserves the large-$N$ 't Hooft limit while at the same time leads to $\mathcal{O}(1)$ anomalous dimensions. If $\Phi$ is multi-trace, the connected correlation functions behave differently depending on the types of states we are trying to lift. In particular, some connected correlation functions scale as  $\mathcal{O}(N^0)$ while most go like $\mathcal{O}(N^{-1})$. It is therefore impossible to lift most states of the theory and to simultaneously preserve the 't Hooft limit with multi-trace deformations (see App.\,D of \cite{Belin:2020nmp} for details).
\end{description}

Combining these three requirements, we find that $\Phi$ should come from a 1/2-BPS state in the twisted sector given by a single cycle of length $n$, or twist-$n$ sector for short. In the end, the actual correlator in \eqref{eq:finalgamma} that we need to evaluate is
\be\label{eq:maincorr}
\bra{W_{s,a}}\hat\Phi(1)\hat\Phi(x)\ket{W_{s,b}}~,
\ee
where $\hat\Phi$ is an operator in the  twist-$n$ sector and $W_{s,a}$ is an untwisted operator. To evaluate this correlator, we want to ``undo'' the twist by using a map that takes fields in Sym$^N(\cal{C})$, the base, to a covering Riemann surface, the cover. Let us first  explain the technicalities of this covering map and how it affects the operators and evaluation of \eqref{eq:maincorr}.

\subsection{Twists and covers}\label{sec:t&c}

In this section we will discuss how to compute correlation functions of twisted fields. The basic idea is to map the correlator to an $n$-sheeted cover \cite{Dixon:1986qv,Lunin:2000yv,Lunin:2001pw,Calabrese:2009qy,Pakman:2009zz,Pakman:2009ab}. 
Fields sitting in the $i$-th tensor factor of the orbifold CFT get mapped to the $i$-th cover sheet, and twisted fields sit at the branch points of the cover map.

In this paper, the correlation functions we are interested in contain two operators in the untwisted sector and two operators in the twist-$n$ sector, the latter of which can be placed at $z=0$ and $z=\infty$ for simplicity. The cover map is then simply given by $z=t^n$, and the cover surface is still a sphere. A twist-$n$ primary field $\phi$ on the base of weight $h$ gets mapped to a field $\tilde \phi$ of weight $\tilde h$ on the cover such that
\be\label{fieldweight}
h = \frac {\tilde h}{ n} +\frac c{24}\left(n-\frac1n\right)\,.
\ee

Next, we discuss what happens to descendant fields. We will closely follow \cite{Lunin:2001pw}.
Let us start with descendants coming from a bosonic chiral field $X$ in the seed theory. Such a field leads to a chiral field in the untwisted sector of the orbifold theory, namely
\be\label{W0}
X^0(z) = {\rm Sym}(X^{(1)}(z) \otimes \mathbb{1} \,\otimes \mathbb{1}\, \dots ) = \sum_{i=1}^N X^{(i)}(z)\,, 
\ee
where $X^{(i)}$ denotes the operator with $X$ in the $i$-th factor and the vacuum in every other factor. By definition, rotating such a field around the insertion of a twist-$n$ field maps $X^{(k)}\to X^{(k+1)}$, from which we see that $X^0$ is indeed invariant under $S_N$. However, besides (\ref{W0}) we can construct a family of operators
\be
X^m(z) \coloneqq \sum_{k=0}^{n-1} X^{(k)}(z) e^{-2\pi i km/n}\ , \quad \quad m \in\{ 0,1,\ldots, n-1\}\ .
\ee
As we rotate around the twist field the new fields pick up phases as in  $X^m \to X^m e^{2\pi i m/n}$. This means that they are no longer integer moded, but rather $1/n$ moded. Since the moding is determined $m$ modulo $n$, we write the modes as $X_{-m/n}$ with the convention that we do not reduce the fraction. We note that of the $X^m(z)$, only $X^0(z)$ is invariant under $S_N$, survives the orbifold projection, and is therefore a bona fide symmetry of the theory with corresponding standard Ward identities. All other $X^m(z)$ need to appear in $S_N$ invariant combinations.

Nonetheless, the $X^m(z)$ fields are crucial to our computations, as they turn into regular descendant modes on the cover: under the cover map $z=t^n$, we have
\be\label{fieldlifting}
\tilde X(t) = X(z)(nt^{n-1})^h\ ,
\ee
so that these modes transform into
\be\label{modelifting}
X_{\alpha}\mapsto n^{1-h}\tilde X_{\alpha n}\ .
\ee

Let us pause to point an important technical issue: even if $\phi$ is a primary on the base, there is no guarantee that $\tilde \phi$ on the cover is also a primary field.\footnote{The fact that primaries of symmetric products can be made out of descendants on the cover was already noticed by \cite{Klemm:1990df,Lunin:2001pw,Headrick:2010zt}.} The usual condition for $\phi$ to be a $X$-primary is simply that it is annihilated by all positive modes of the only bona fide current $X^0(z)$. However, as we can see, on the cover that only guarantees that it will be annihilated by modes of the form $\tilde X_{kn}$. To guarantee that $\tilde \phi$ is annihilated by  all positive modes of $\tilde X$, we want to impose that $\phi$ is annihilated also by all positive fractional modes of all the $X^m(z)$. In practice this means we want to write $\phi$ as a fractional descendant of such a genuine primary field. This guarantees that $\tilde\phi$ will be a genuine primary field on the cover. The fractional descendants can then be taken care of by (\ref{modelifting}).

Next, we turn to fermionic operators, where the story is slightly more subtle. Even though we will always work in the NS sector on the base, fermionic operators may need to end up in the Ramond sector on the cover. To see this, let us investigate what happens to $\tilde X(t)$ as we rotate around 0 by sending $t\to te^{2\pi i}$. On the base this means we rotate $n$ times around the twist field, so that we indeed end up on the same sheet. For bosonic operators with integer $h$, the additional factor in (\ref{fieldlifting}) is simply 1. We find that $\tilde X(t)$ gets sent to itself, which is exactly what we expect. For $h$ half-integer however, we find that the factor gives a sign $(-1)^{n-1}$. Concretely, this means that if on the base we are in the NS sector so that $X(z)$ does not pick up a sign under rotation, $\tilde X(t)$ picks up a sign $(-1)^{n-1}$ when rotating around $\tilde \phi$. If $n$ is even, $\tilde \phi$ is therefore in the Ramond sector.

We can now work out what this means for $G(z)$ in the NS sector on the base. Because $G$ has half-integer modes, the modes in the expansion read
\be
G^m(z) = \sum_{\alpha=1/2+m/n+\Z} G^m_{\alpha} z^{-\alpha-3/2}\,.
\ee
As expected, we see that cover modes $\tilde G_{\alpha n}$ are half-integer if $n$ is odd, and integer when $n$ is even.

A further remark on the modes of $G$ is in order.
If $n$ is even, then $\tilde G(t)$ has a zero mode. That zero mode  comes from the field
\be
G^{n/2}(z) = \sum_{k=0}^{n-1}(-1)^k G^{(k)}(z)
\ee
on the base. Even though we are in the NS sector, due to the twist this field picks up a minus sign under rotation around the origin, and is therefore integer moded. For concreteness, let us pick $\phi$ to be a chiral primary field that is annihilated by all positive fractional modes of weight 1/2 and charge 1, so that $G^-_{-1/2}\phi$ is a modulus. $G^{+,n/2}_0$ then increases the charge by 1 and leaves the weight invariant. Since $G^{+,n/2}_0\phi$ violates the unitarity bound, we necessarily must have $G^{+,n/2}_0\phi=0$. On the other hand $|G^{-,n/2}_0\phi|^2=n(1/2-c/24)$, so as long as $c<12$ this state is non-vanishing. In conclusion this means that on the cover we should have
\be
G^+_0 \tilde \phi=0 \ , \qquad G^-_0\tilde \phi \neq 0\ .
\ee

Let us briefly mention the implications for the twisted sector ground states. If $n$ is odd, the twist-$n$ ground state $\sigma_n$, that is the lightest state in the twist-$n$ sector, will get mapped to the NS vacuum on the cover. Since $\tilde h=0$, we find that its weight on the base equals
\begin{align}\label{gsoddn}
    h_{\sigma_n}= \frac{c}{24}\left(n-\frac{1}{n}\right)\,.
\end{align}
If $n$ is even, then NS sector states on the base are mapped to the Ramond sector on the cover. The lightest Ramond sector state on the cover has $\tilde h =\frac c{24}$, which corresponds to
\be\label{gsevenn}
h = \frac {cn}{24}\ .
\ee
The difference between (\ref{gsoddn}) and (\ref{gsevenn}) is usually attributed to a spin field acting on $\sigma_n$.

\subsection{Moduli}\label{sec:moduli}

Let us now discuss how to find moduli $\Phi$ that satisfy the three properties listed at the beginning of Sec.~\ref{sec:coverbasemap}. We will see that the existence of such moduli already imposes conditions on the seed theory ${\cal C}$ of Sym$^N({\cal C})$.   

First, the modulus is a $G_{-1/2}$ descendant of a of a 1/2-BPS supersymmetric operator, that is a chiral or anti-chiral primary in the NS sector as defined in \eqref{eq:chiralmariginal}. For concreteness, in the following we will focus on the chiral case and focus on the left-moving (holomorphic) pieces of the operator. We write 
\begin{align}\label{eq:marginalstate}
    G^{-}_{-1/2}|\varphi_c\rangle \,,
\end{align}
where $\varphi_c$ is a chiral primary with $h_{\varphi}=1/2$ and $q=1\,$.

Second, because the modulus is in the twist-$n$ sector, schematically the chiral primary is of the form
\begin{align}\label{eq:def-Phi}
   \ket{\varphi_c} = \phi_{(n)} \ket{\sigma_n} \,,
\end{align}
where $\ket{\sigma_n}$ is the ground state of the twist-$n$ sector with weight given by (\ref{gsoddn}).
In \eqref{eq:def-Phi}, $\phi_{(n)}$ is an appropriate combination of operators such that $\varphi_c$ is a 1/2-BPS operator with $h_{\varphi}=1/2$ and $q=1\,$. In particular, in view of the discussion of the Ramond sector above, for even $n$ it will contain what is usually called the spin field.
Under the cover map, this $\phi_{(n)}$ will get mapped to some field $\tilde \phi$ on the cover. Let us derive what properties it has.

To start, we take $h_{\varphi}=1/2$ and plug it into (\ref{fieldweight}) to recover the following expression for the weight $\tilde h_\phi$ of $\tilde\phi$ on the cover,
\begin{align}\label{eq:hphionthecover}
   \tilde h_{\phi}= \frac{n}{2}-\frac{c}{24}\left(n^2-1\right)\,.
\end{align}
For \textbf{odd} twist, we are in the NS sector on the cover, and the unitarity bound is simply that $\tilde h_{\phi}\geq |q|/2=1/2$, since going to the cover leaves the charge invariant. 
This leads to the restriction
\begin{align}\label{eq:cnodd}
    c(n+1)\le 12 \,.
\end{align}
For \textbf{even} twist, we are in the Ramond sector, which means that $\tilde \varphi$ cannot be lighter than the Ramond ground state, \ie $\tilde h_{\phi}\geq \frac c{24}$. This leads to the restriction
\begin{align}\label{eq:cneven}
    c\,n\leq 12\,.
\end{align}
These two inequalities should be viewed as restrictions on possible values of $c$ and $n$. In particular, for each discrete fixed value of $n$ we can easily determine the highest possible value of the central charge of ${\cal C}$. The resulting values are stated in Table \ref{tab:cmax}. 
As expected \cite{Belin:2020nmp}, it is impossible to find a modulus of the required properties if $c$ is greater than 6. It is interesting to see however that even if $c$ falls between $1\leq c \leq6$, only certain specific values of $n$ are allowed.

\begin{table}[ht!]\def\arraystretch{1.5}
\centering
\begin{tabular}{|c|c|c|c|c|c|c|c|c|c|c|c|}
\hline
     $n$&2&3&4&5&6&7&8&9&10&11&12  \\
     \hline
     $c_{\rm max}$& 6&3&3&2&2&$\frac{3}{2}$&$\frac{3}{2}$&$\frac{6}{5}$&$\frac{6}{5}$&1&1 \\
\hline     
\end{tabular}
\caption{Upper bound $c_{\rm max}$ on the central charge of the seed theory for which a twist $n$ marginal operator can exist.}\label{tab:cmax}
\end{table}

We will discuss the composition of $\tilde \phi$ in detail for specific seed theories in the next section. But before that, let us point out some interesting generalities based on the information so far:
\begin{itemize}
    \item For $n=2$, we see from \eqref{eq:hphionthecover} that
    \begin{align}\label{eq:n2general}
        \tilde{h}_{\phi_{(2)}}-\frac{c}{24}= 1-\frac{c}{6}~.
    \end{align}
    Note that we have subtracted the energy arising from the Ramond ground state. For $c=6$ the state is trivial, as it is well known. For $c<6$, the state $\tilde \phi$ has to be a primary in the R-sector on the cover with $U(1)$ charge $q=1$. 
    \item For $n=3$, we have
     \be\label{eq:n3general}
\tilde{h}_{\phi_{(3)}}= \frac{1}{2}+\left(1-\frac{c}{3}\right)~.
\ee
We explicitly split the result into two parts as this suggests that the state $\tilde{ \phi}_{(3)}$ is a $G^{+}_{-1/2}$ descendant of a neutral primary on the cover with $h= 1-\frac{c}{3}$. It turns out that this is always the case.\footnote{Note that for $c=3$ the state $\tilde{ \phi}_{(3)}$ has $\tilde{h}_{\phi_{(3)}}= \frac{1}{2}$; this is compatible with the twist-3 operator discussed in \cite{Gukov:2004ym,Elitzur:1998mm} for symmetric product theories dual to AdS$_3\times S^3\times S^3\times S^1$.}
    \item The constraints \eqref{eq:cneven} and \eqref{eq:cnodd} are necessary conditions for a twisted marginal operator to exist, but they are by no means sufficient. For example, the bound in \eqref{eq:cneven} does not incorporate the conditions that the $U(1)$ charge of $\phi_{(n)}$ is $q=1$. The values in the table listed above, although stringent, will be further constrained as we revisit the moduli for seed theories with $c<3$, i.e. for the minimal models.
\end{itemize}

Finally we want to briefly remark that $\varphi$ is technically not a moduli yet, but needs to be symmetrized as in (\ref{prestate1}). In particular this introduces a $N$-dependent normalization factor. We discuss normalizations in more detail in App.\,\ref{app:largeN}, and simply remark that this part of the normalization actually drops out in the final computation of the correlation function.

\subsection{\texorpdfstring{$\mathcal{N}=2$}{N=2} minimal models and their twisted moduli}\label{sec:mmmoduli}

In this section we will focus on the case when the seed theory is a ${\cal N}=2$ minimal model. For these theories, all possible marginal operators, including untwisted and twisted sectors, were singled out in   \cite{Belin:2020nmp}. Here we will revisit this result, focusing only on the single-trace moduli, and decode in detail the composition of $\tilde \phi$ from the perspective of the cover space. 

$\mathcal{N}=2$ minimal models are unitary SCFTs, and constitute a complete classification of unitary CFTs with $\mathcal{N}=(2,2)$ supersymmetry and $c<3$ \cite{Boucher:1986bh, DiVecchia:1986fwg}. They come labeled by a positive integer $k$ related to the central charge by
\be\label{eq:cmm}
c = \frac{3k}{k+2}~.
\ee
The classification of minimal models is naturally related to simply-laced Dynkin diagrams, which define the $A$, $D$, and $E$ series. Each of these families are labelled as
\begin{itemize}
\item the $A$-series, which have $c=\frac{3k}{k+2}$ for any positive integer $k$ and are denoted $A_{k+1}$;
\item the $D$-series, which have $c=\frac{3k}{k+2}$ for any even $k \geq 4$ and are denoted $D_{k/2+2}$; 
\item and three exceptional theories denoted $E_6$, $E_7$, and $E_8$, which have $c=\frac 52, \frac 83$, and $\frac{14}5$, respectively. 
\end{itemize}
A defining property of minimal models is that they are rational CFTs: they contain a finite number of irreducible representations of the superconformal algebra. We will denote the highest weight state of these representations as   
\be\label{eq:hws}
\varphi^{\rm R}_{(r,s)}~,\qquad  \varphi^{\rm NS}_{(r,s)}~,
\ee
where we are making an explicit distinction between the Ramond and Neveu-Schwarz sectors of the theory. The labels $(r,s)$, are explained in App.\,\ref{app:minmodel}, where we also write their conformal dimension and $U(1)$ charge.

One of the results in \cite{Belin:2020nmp} was to explicitly identify the possible supersymmetric marginal operators in the single-trace twisted sector of Sym$^N({\cal C})$ when ${\cal C}$ is an $\mathcal{N}=2$ minimal model.\footnote{The analysis of \cite{Belin:2020nmp} also includes  multi-trace twisted and untwisted moduli, but they are not the focus of the work here.} The summary of that analysis is stated in Table \ref{t:moduli}. Notice that not all allowed values of the twist we infer from \eqref{eq:cneven} and \eqref{eq:cnodd} are listed in Table \ref{t:moduli}. This reflects the fact that for $c<3$ the precise spectrum of the seed further restricts $n$. If we combine the results of Table \ref{t:moduli} with \eqref{eq:cneven} and \eqref{eq:cnodd}, we conclude that the only allowed values of the twist for a supersymmetric marginal operator are 
\be
n=\{2,3,4,5,7\}\,,
\ee
and those are the cases we will consider hereafter.  

\begin{table}[ht!]
	\centering
	\begin{tabular}{|ccc|}
		\hline
		Series & $k$  & moduli\\
		\hline
		$A_2$ & 1&1 twist 5, 1 twist 7\\
		$A_3$ & 2 &1 twist 3, 1 twist 4, 1 twist 5\\
		$A_5$ & 4 &1 twist 2, 1 twist 3, 1 twist 4\\
		$A_{k+1}$ & odd, $\geq3$ & 1 twist 3\\
		$A_{k+1}$ & even, $\geq6$ &1 twist 2, 1 twist 3\\
		\hline
		$D_4$ & 4 & 1 twist 2, 2 twist 3, 1 twist 4\\
		$D_{\frac{k}{2}+2}$ & $0~\text{mod}~4, ~\geq 8$ &1 twist 2, 1 twist 3\\ 
		$D_{\frac{k}{2}+2}$ & $2~\text{mod}~4, ~\geq 6$ &1 twist 3\\
		\hline
		$E_6$ & 10 & 1 twist 2\\
		$E_7$ & 16 & 1 twist 2 \\
		$E_8$ & 28 & 1 twist 2 \\
		\hline
	\end{tabular}
	\caption{Number of twisted sector single-trace moduli  for symmetric product orbifolds of the $ADE$ minimal models. The central charge of the minimal models is related to the parameter $k$ by $c=\frac{3k}{k+2}$.}
	\label{t:moduli}
\end{table}

In \cite{Belin:2020nmp} the existence of these marginal operators was established by reading off terms from the orbifold partition function. Here we need to construct them explicitly by finding a suitable state on the cover with weight \eqref{eq:hphionthecover} and charge $q=1$. We express $\tilde \phi$ in terms of  \eqref{eq:hws} on the covering space, which turns out to be more intricate than expected.   

We start by first describing $\tilde \phi_{(n)}$ for $n=2$. As we remarked in \eqref{eq:n2general}, the state on the cover has to be a primary with
 \be
\tilde{h}_{\phi_{(2)}}-\frac{c}{24}= \frac{1}{2}+\frac{1}{k+2}~,
\ee
where we used \eqref{eq:cmm}. Matching this condition with the allowed highest weight representations of the minimal models gives
\be\label{eq:modulin2}
 \tilde \phi_{(2)} = \varphi^{\rm R}_{(r,s)} ~, \quad {\rm with} \quad r= \frac{k}{2}+3 ~,\quad s= \frac{k}{2}+1~,
\ee
and moreover $k$ is even and $k\geq4$. Although the primary $\varphi^{\rm R}_{(r,s)}$ identified on the cover  is not the same as the one used on the base, we emphasize that there is no tension: the allowed values of $k$ perfectly agree with Table \ref{t:moduli}, and when the state \eqref{eq:modulin2} is mapped back to the base it leads to the state reported in \cite{Belin:2020nmp}.  This will be the case in all subsequent states analysed below. 

 For $n=3$ the results can be described rather easily as well, but there are some important differences relative to $n=2$. First since $n$ is odd, the boundary conditions of the fermions are unchanged, and on the cover we have  
 \be
\tilde{h}_{\phi_{(3)}}= \frac{1}{2}+ \frac{2}{k+2}~,
\ee
where we inserted \eqref{eq:cmm} in \eqref{eq:n3general}. Matching this condition with the allowed highest weight representations of the minimal models gives
\be\label{eq:modulin3}
 \tilde \phi_{(3)} = G^{+}_{-1/2}\varphi^{\rm NS}_{(r,s)} ~, \quad {\rm with} \quad r= 3 ~,\quad s= -1~,
\ee
which requires that $k\geq2$. This gives the second difference relative to $n=2$: the state $ \tilde \phi_{(3)}$ on the cover is a descendant of a primary. Again there is agreement with \cite{Belin:2020nmp}.  

The remaining values of $n$ are very sporadic in Table \ref{t:moduli}, and the final outcome is the following. There are three special cases:
\begin{equation}
    \begin{aligned}
    &k=1\,,~ n=5: \quad \tilde \phi_{(5)} = G^{+}_{-3/2}\vac^{\rm NS}~,\\
    &k=1\,, ~n=7: \quad \tilde \phi_{(7)} = G^{+}_{-3/2}\vac^{\rm NS}~,\\
    &k=2\,, ~n=4: \quad \tilde \phi_{(4)} = G^{+}_{-1}\vac^{\rm R}~.
    \end{aligned}
\end{equation}
It is rather surprising that for these cases the state on the cover is simply a descendant of the vacuum state: for very low values of $c$ the existence of a primary is guaranteed just by the covering map, and does not depend on the seed spectrum at all. The remaining two cases are
\begin{equation}
    \begin{aligned}
    &k=2\,,~ n=5: \qquad \tilde \phi_{(5)} = G^{+}_{-1/2} \varphi^{\rm NS}_{(3,-1)}~,\\
    &k=4\,, ~n=4: \qquad \tilde \phi_{(4)} = \varphi^{\rm R}_{(5,3)}~.
    \end{aligned}
\end{equation}
Note that these expressions follow the trends for twist 2 and 3: for $k=2$, $n=5$ the state is of the form \eqref{eq:modulin3} while for $k=4$, $n=4$ the state complies with  \eqref{eq:modulin2}. 

\bigskip

To summarize: the composition of $\tilde\phi_{(n)}$ on the cover is sensitive to $n$, $c$, and the spectrum of the seed theory $\mathcal C$. In particular, while the field $\phi_{(n)}$ is always an $\mathcal N = 2$ primary on the base, the field $\tilde\phi_{(n)}$ can be a primary or a descendant. This is important since correlation functions of twist fields are evaluated on the cover space using the $\mathcal N = 2$ Ward identities, the latter of which differ for primary and descendant fields. As a result, the anomalous dimensions of the higher spin currents of ${\rm Sym}^N({\cal C})$ are not generically universal but depend on the details of the seed theory $\mathcal C$.

 \subsection{Correlation functions}\label{sec:cfs}
 
 Now that we have understood the structure of both the currents and the marginal operators, we can shift our focus to the evaluation of the correlator appearing in (\ref{eq:finalgamma}), namely
\be
 \bra{W_{s,a}}\big(G^+_{-1/2}\varphi_a\big)(1)\big(G^-_{-1/2}\varphi_c\big)(x)\ket{W_{s,b}}\, .\label{eq:basecf1}
 \ee
There is a similar term with the two moduli interchanged which can be evaluated in a similar way.

First of all, we can simplify the computation by permuting the operators so that the twist insertions are located at $z=0$ and $z=\infty$. This will simplify the covering map considerably, as was noted in Sec.\,\ref{sec:t&c}. The permutation of the operators can be achieved by adding an appropriate regulator
 \begin{align}
     \begin{split}
& \bra{W_{s,a}}\big(G^+_{-1/2}\varphi_a\big)(1)\big(G^-_{-1/2}\varphi_c\big)(x)\ket{W_{s,b}}\\
 &\hspace{30pt}=\lim_{\delta\rightarrow0} \delta^{-2s}\left\langle W_{s,a}(\delta^{-1})\big(G^+_{-1/2}\varphi_a\big)(1)\big(G^-_{-1/2}\varphi_c\big)(x)W_{s,b}(0)\right\rangle\, ,\label{eq:reg}
 \end{split}
 \end{align}
 and performing the following coordinate transformation, that sends a complex coordinate $z$ to
  \be
 z\mapsto f(z)\coloneqq\frac{z-x+(x-1)\delta}{z-1-(x-1)\epsilon}\,.\label{eq:permutation}
 \ee
This transformation maps the insertion points of \eqref{eq:reg} to 
\eq{
0\mapsto x\,,\quad 1\mapsto \epsilon^{-1}\,,\quad x\mapsto 0\,,\quad \delta^{-1}\mapsto 1\,.
}
We let $\epsilon$ be a regulator to define a normalized bra state again. By using \eqref{eq:permutation}, we end up, to leading order in $\epsilon$ and $\delta$, with
 \begin{align}\label{eq:11}
     \begin{split}
&\bra{W_{s,a}}\big(G^+_{-1/2}\varphi_a\big)(1)\big(G^-_{-1/2}\varphi_c\big)(x)\ket{W_{s,b}}\\
&\hspace{120pt}=(x-1)^{2s-2}\bra{\varphi_c}G^+_{1/2}W_{s,a}(1)W_{s,b}(x)G^-_{-1/2}\ket{\varphi_c}\, .
\end{split}
\end{align}

Next, we expand the fields in terms of their constituents, i.e.~in terms of the operators on the $i$-th copy of Sym$^N({\cal C})$. Starting with the currents $W_{s,a}$ and $W_{s,b}$, the four-point function \eqref{eq:11} is a linear combination of correlation functions of the form
\be
\bra{\varphi_c}G^+_{1/2}\big(\prod_{k=1}^{m_1} \partial^{a_k}O_k^{(j_k)}\big)(1)\big(\prod_{\ell=1}^{m_2} \partial^{a'_\ell}O_\ell^{(j'_\ell)}\big)(x)G^-_{-1/2}\ket{\varphi_c}\, ,\label{eq:basecf}
\ee
 where $O_k^{(i)}\in\{J^{(i)},G^{\pm(i)},T^{(i)}\}\,$, $a_k\in\mathbb{Z}_{\ge0}\,$, and the respective weights $h_k$ add up to the spin of the current
\be
\sum_{k=1}^{m}h_k=s\,.
\ee
In order to reconstruct \eqref{eq:11}, we have to sum over the distinct $j_k$'s. Note that this is the step where we can utilize the freedom to write the currents in ways that either highlights their global terms or their single and multi-trace terms; see, for example, the contrast between \eqref{W4} and \eqref{W42}. 

We will now start by giving the correlation function that has to be computed on the cover, and go through the steps needed to find the base correlator \eqref{eq:basecf}. On the cover we must compute
\be
\bra{\tilde\phi_{(n)}}\tilde G^+_{n/2}\big(\prod_{k=1}^{m_1} \tilde \partial^{a_k}O_k^{}(t_k)\big)\big(\prod_{\ell=1}^{m_2} \tilde \partial^{a'_\ell}O_\ell^{}(t_{m_1+\ell})\big)\tilde G^-_{-n/2}\ket{\tilde\phi_{(n)}}\, .
\ee
Note that this correlator only contains two $\cN=2$ primaries dressed with $G^\pm$ modes and insertions of (derivatives of) $J$, $G^\pm$ and $T$. Therefore, it can be completely evaluated using just the $\cN=2$ Ward identities (see App.\,\ref{app:WI} for a review). In order to get back the base correlator, we then have to carry out the following steps.

First, we have to implement the cover map $z=t^n$, or more precisely, the inverse of this map, to map the correlator on the cover to a correlator on the base. The inverse of the cover map is given by
    \be
    t=
    z^{1/n}\zeta_n^k\, ,
    \ee
    where $k$ denotes the copy on the base that $t$ lands on under this map, and $\zeta_n$ is defined as a root of unity of order $n$
    \[
    \zeta_n\coloneqq e^{2\pi i/n}\,.
    \]
    Furthermore, we have to incorporate the transformation of the operators under the map. Since $J$ and $G^{\pm}$ transform as Virasoro primaries under conformal transformations we have
    \be
    J^{(k)}(z)=\left(\frac{\partial z}{\partial t}\right)^{-1}\tilde J(t)=
    \frac{z^{1/n}\zeta_n^k}{nz}\tilde J(t)\,,
    \ee
    and
    \be
    \left(G^\pm\right)^{(k)}(z)=\left(\frac{\partial z}{\partial t}\right)^{-3/2}\tilde G^\pm(t)=
    \left(\frac{z^{1/n}\zeta_n^k}{nz}\right)^{3/2}\tilde G^\pm(t)\,.
    \ee    
    The stress tensor, however, transforms with an additional contribution from the Schwarzian derivative
    \begin{align}
    \begin{split}
    T^{(k)}(z)&=\left(\frac{\partial z}{\partial t}\right)^{-2}\left(\tilde T(t)-\frac{c}{12}\{z,t\}\right)\\
    &=\left(\frac{z^{1/n}\zeta_n^k}{nz}\right)^2\left(\tilde T(t)+\frac{c}{24}\frac{n^2-1}{t^2}\right)\,.
    \end{split}
    \end{align}
    Moreover, using \eqref{modelifting} we see that the $\tilde G^\pm_{n/2}$ modes that dress the twist operator are mapped as follows
    \be
    \tilde G^{\pm}_{n/2}\mapsto n^{1/2}G^{\pm}_{1/2}\,,
    \ee
    so that $\tilde G^{-}_{n/2}\ket{\tilde{\phi}_{(n)}}$ is indeed mapped to $G^-_{-1/2}\ket{\varphi_c}$ (dressed with a factor of $\sqrt{n}$). The result of implementing the covering map is that we now have computed the following correlator on the base
    \be
\bra{\varphi_c}G^+_{1/2}\big(\prod_{k=1}^{m_1} \partial^{a_k}O_k^{(j_k)}(z_1)\big)\big(\prod_{\ell=1}^{m_2} \partial^{a'_\ell}O_\ell^{(j'_\ell)}(z_2)\big)G^-_{-1/2}\ket{\varphi_c}\, ,\label{eq:basecor1}
    \ee
    where we will take $z_1=1$ and $z_2=x$.

Second, we may need to normal order (some of) the operator insertions in \eqref{eq:basecor1}. Normal ordering is necessary when we have $k_1,k_2\in\{1,\dots,m_1\}$ such that $j_{k_1}=j_{k_2}$ (and similarly for equal $j_\ell$'s). Explicitly, we use
\be
(O_1O_2)^{(j)}(z)=\frac{1}{2\pi i}\oint_z\frac{dw}{w-z}O_1^{(j)}(w)O_2^{(j)}(z)\, .
\ee
An important detail here is that we should consider all the operators that we want to normal order on the base. If we had done the normal ordering procedure on the cover, which may naively seem simpler, the composite operator is generically 
not a Virasoro primary. This implies that the transformation under the inverse cover map is considerably more complicated, which is something that we would like to avoid.
 
Finally, we need to take into account the fact that all fields are symmetrized, which means that we need to sum over $S_N$ and also take into account the proper normalization of states, as was explained in Sec.\,\ref{sec:largeN}. Luckily, this can be simplified considerably, as is explained in detail in App.\,\ref{app:largeN}: the twisted sector moduli can be fixed to be in the first $n$ copies. If both the operators at 1 and $x$ are single-trace, then the sums over $j_1$ and $j'_1$ run from 1 to $n$ only, since any other configuration leads to a disconnected (or even vanishing) contribution, which cannot give a logarithmic term.

For multi-trace operators the situation is slightly more complicated. At least one of the $j_k$ still needs to be in the first $n$ factors, but the remaining $j_k$ and $j'_k$ can be in the other $N-n$ copies. It follows that the higher the trace, the more sums over the large range $N-n$ we may have to compute. The first non-trivial case that we will encounter is the currents having a double-trace term. In that scenario we will have to compute the following terms 
 \begin{align}
 \begin{split}
   &\sum_{\substack{j_1,j_2=1\\j_1\neq j_2}}^N\sum_{\substack{j_3,j_4=1\\j_3\neq j_4}}^N\bra{\varphi_c}G_{1/2}^+O^{(j_1)}(1)O^{(j_2)}(1)O^{(j_3)}(x)O^{(j_4)}(x)G^-_{-1/2}\ket{\varphi_c}\\
    &\hspace{20pt}=\sum_{\substack{j_1,j_2=1\\j_1\neq j_2}}^n\sum_{\substack{j_3,j_4=1\\j_3\neq j_4}}^n\bra{\varphi_c}G_{1/2}^+O^{(j_1)}(1)O^{(j_2)}(1)O^{(j_3)}(x)O^{(j_4)}(x)G^-_{-1/2}\ket{\varphi_c}\\
    &\hspace{40pt}+\sum_{j_1,j_3=1}^n\sum_{j_2=n+1}^N\bra{\varphi_c}G_{1/2}^+O^{(j_1)}(1)O^{(j_3)}(x)G^-_{-1/2}\ket{\varphi_c}\left\langle O^{(j_2)}(1)O^{(j_2)}(x)\right\rangle\\
    &\hspace{40pt}+\ \text{permutations}\,.
\end{split}    
 \end{align}
One thing to note here is that since the twist is constricted to the first $n$ copies, the operators that appear in the disconnected part of the correlator must be on the same copy. This is because the only reason a correlation function between untwisted operators on different copies can be non-zero is the presence of a twisted operator that allows the different sheets to communicate with one another. Moreover, because the disconnected part has only untwisted operators in it, the sum over the final $N-n$ copies simplifies to just a seed correlation function, namely
\be
\sum_{j=n+1}^N\left\langle O^{(j)}(1)O^{(j)}(x)\right\rangle=(N-n)\left\langle O(1)O(x)\right\rangle\, .
\ee

Evaluating the sums over the copies with the twist insertions is more involved. In terms of the correlator on the covering space it boils down to a sum over insertions at roots of unity, which is generally hard to compute. At leading order in $N$, when the currents are single-trace, we can evaluate the sums analytically (see App.\,\ref{sec:sums} for a proof of this statement). However, in order to compute the finite $N$ result, we must also compute terms involving multi-trace insertions, and thus more sums. Even though these sums are generically difficult to deal with, we are saved by the fact that the twists we consider are not that high. From Table \ref{t:moduli} we can read off that at worst we have to sum roots of unity of order $n=7$. This number is low enough that we are able to compute the sums explicitly.

By carrying out the steps described above we can evaluate the correlation function \eqref{eq:basecf}. Using the regularization procedure described in Sec.\,\ref{sec:i&r} we now have all the ingredients necessary to compute the anomalous dimensions of higher spin currents described in the following section. We used Mathematica for our computations and the notebooks are available upon request.

\section{Lifting currents}\label{sec:lifths}

We are now ready to compute the anomalous dimensions of the first currents, up to spin three, that will lift under the deformation of the symmetric orbifold. The discussion in this section incorporates all the ingredients that a generic ${\cal N}=2$ SCFT seed theory brings to this analysis. More prominently, our analysis incorporates all the intricacies and novelties of the moduli laid out in Sec.\,\ref{sec:coverbasemap}, which is not universal as a function of $c$, nor the twist.  We also discuss the large-$N$ behaviour of our results and contrast them with existing results.  

We stress that the currents we consider in this paper are those that are built from the $\cN=2$ superconformal currents. In the case that the seed theory $\cC$ has additional currents, the latter have to be incorporated by following the steps laid out in Sec.\,\ref{sec:currents}. These additional currents would enlarge the mixing matrix $\gamma_{ab}$, and hence potentially affect the magnitude of the eigenvalues $\mu_a$ reported here. Therefore, our expressions are only strictly valid when there are no such additional currents present. This is the case, for example, for the (infinite) family of A-series minimal models. In general, we expect that all higher spin currents will lift even in cases with additional currents, albeit this needs to be confirmed explicitly.

\subsection{Spin-2 sector} \label{se:spin2sector}

We start with the construction of the appropriate current. In Sec.\,\ref{sec:currents}, we already saw that any symmetric orbifold contains a new primary current at spin four built entirely from the stress-tensor. It is then not surprising that any symmetric product theory containing a stress tensor and a $U(1)$ current (such as any $\mathcal{N}=(2,2)$ CFT)  will contain a new primary spin-two current. There are two linearly-independent Virasoro primaries at weight 2 that are given by:
\begin{align}
 T(z) - \frac{3}{2} \sum_{i = 1}^N (J^{(i)} J^{(i)})(z)~, \qquad  \sum_{i \ne j}^N J^{(i)} (z) J^{(j)} (z)\,. \label{W2X2def}    
\end{align}
The OPE of these operators with $T(z)$ reflect that they are primaries with respect to the Virasoro algebra, however they are not primaries of the Virasoro-Kac-Moody algebra, i.e.~the algebra generated by the stress tensor $T(z)$ and the $U(1)$ $R$-current $J(z)$. One combination that will be pedagogical to consider is given by 
\begin{align}
Y_2(z) = T(z) - \frac{3}{2}(JJ)(z) \,. \label{Y2def}    
\end{align}
It is important to note that $Y_2$ is built out of the full stress tensor and $U(1)$ current of the theory. Hence, since the marginal deformation preserves the $\mathcal{N}=(2,2)$ superconformal symmetry, the conformal dimension of $Y_2$ is protected on the conformal manifold. Nevertheless, we will evaluate the appropriate correlation functions and integrals in \eqref{eq:exp}  to stress  that it does not acquire an anomalous dimension. 

The linear combination of \eqref{W2X2def} that leads to the non-trivial current at spin-two is
\begin{align}
W_2(z) &= T(z) - \frac{3}{2}(JJ)(z) + \frac{3(c N - 1)}{2c(N -1)}\sum_{i \ne j}^N J^{(i)}(z) J^{(j)}(z) \label{W2def}
\,.   
\end{align}
This operator is a Virasoro-Kac-Moody primary, and a primary under the full ${\cal N}=2$ algebra. As we will report, it is not protected under the marginal deformation. In particular, note that the double-trace components of $W_2$ are subleading in $N$ such that in the large-$N$ limit, the spin-2 current consists only of single-trace operators, namely 
\begin{align}
W_2(z) = \sum_{i = 1}^N \big[ T^{(i)}(z) - \frac{3}{2} (J^{(i)}J^{(i)})(z)\big] + \mathcal O(1/N)~.    
\end{align}

We will explicitly evaluate the anomalous dimensions of both $(Y_2, W_2)$ although only that of $W_2$ can be nonzero. First, we normalize the operators in accordance to \eqref{eq:deformation}. Using \eqref{W2X2def}, \eqref{Y2def}, and the Virasoro-Kac-Moody Ward identities, we find
\eq{
 \Vev{Y_2 | Y_2} &=   \frac{c N (c N -1)}{2} \,, \qquad \Vev{W_2 | W_2} &=  \frac{N(c-1)(cN -1)}{2(N-1)} \,, \qquad \Vev{Y_2 | W_2} &= 0\,.
}
It is interesting to note that the $W_2$ field becomes null when $c = 1$, in which case it is not part of the spectrum. For any other value of $c$, the normalized fields are given by an overall re-scaling such that
\begin{align}
  Y_2 \to \sqrt{\frac{2}{c N (c N-1)} } \,Y_2\,, \qquad W_2 \to \sqrt{\frac{2(N-1)}{N(c-1)(c N -1)}}\,W_2 \,.   \label{normW2}
\end{align}

Using the normalization \eqref{normW2}, and following the procedure described in Sec.\,\ref{sec:coverbasemap}, we find that the anomalous dimensions of the spin-2 fields $(Y_2, W_2)$ can be written as
\eq{
\gamma_{(2)} = \left( \begin{array}{cc}
                    \,0 & 0\! \\
                    \,0 & \mu_{(2)}\!{}
                    \end{array} \right). \label{spin2matrix}
}
We observe that the anomalous dimension of $Y_2$ vanishes exactly, as expected, and that the only nonvanishing entry in \eqref{spin2matrix} is the anomalous dimension of $W_2$. This is consistent with the fact that the conformal dimension of $Y_2$ is protected on the conformal manifold. The anomalous dimensions $\mu_{(2)}$ of $W_2$ for the symmetric orbifolds of $\mathcal N = 2$ minimal models are given in Table \ref{spin2table}. It is worth remarking that for this simple case the $N$-dependence of the anomalous dimensions comes solely from the normalization of the $W_2$ current \eqref{normW2}. 
\begin{table}[ht!]
  \begin{center}\def\arraystretch{1.65}
  \begin{tabular}{|c|c|p{1cm}<{\centering}|p{3.80cm}<{\centering}|}
\hline
  $k$ & $c = \frac{3k}{k+2} $ & $n$ & $\mu_{(2)}$ \\ 
  \hline 
  \multirow{2}{*}{1} &  \multirow{2}{*}{1} & 5 &    \multirow{2}{*}{---}  \\
  \cline{3-3}
  &  & 7 &   \\
 \hline
   \multirow{3}{*}{2} &  \multirow{3}{*}{$\frac{3}{2}$} & 3 & $\frac{20 \pi^2 \lambda^2 (3N - 2)}{27( N - 1)}$ \\
  \cline{3-4}
  &  & 4 & $\frac{187 \pi^2 \lambda^2 (3N - 2)}{256( N - 1)}$   \\
    \cline{3-4}
  &  & 5 & $\frac{4 \pi^2 \lambda^2 (3N-2) }{5(N-1)}$  \\
   \hline
 3 & $\frac{9}{5}$ & 3 & $\frac{44 \pi^2 \lambda^2 (9N - 5)}{243( N - 1)}$\\
 \hline
   \multirow{3}{*}{4} &  \multirow{3}{*}{2} & 2 &  $\frac{39 \pi^2 \lambda^2 (2 N - 1)}{64 (N - 1)}$  \\
  \cline{3-4}
   &  & 3 & $\frac{19\pi^2\lambda^2 (2 N -1)}{27(N-1)}$  \\
   \cline{3-4}
  &  & 4 &  $\frac{207 \pi^2 \lambda^2 (2N - 1)}{256( N - 1)}$  \\
  \hline
  $5,6, \dots$ & $2 < c < 3$ & 3 & $\frac{4 \pi^2 \lambda^2 ( c^2 + 12 c -9 ) (c N - 1)}{27c^2 (c - 1)(N - 1)}$ \\
 \hline
  $6,8,\dots$ & $2 < c < 3$ & 2 & $\frac{3 \pi^2 \lambda^2 (24 + c)  (c N - 1)}{64c (c - 1)(N - 1)}$\\
 \hline
  \end{tabular}
  \caption{The anomalous dimensions of the spin-2 current $W_2$ of the symmetric orbifolds of the $\mathcal N = 2$ minimal models. When $c = 1$, the $W_2$ current becomes null and the anomalous dimension of the unnormalized current vanishes. Note that for $n=3$, the anomalous dimension is strictly greater than zero for all $\frac{3}{2}\leq c < 3$.}
  \label{spin2table}
  \end{center}
  \end{table}
  
Note that, as mentioned above, the $W_2$ current becomes null when $c = 1$ and is not a part of the spectrum. In this case, we expect the anomalous dimension to vanish, since the correlation functions of a null state against any other physical state of the theory should be zero. This is evident for the twist $n = 5, 7$ entries with $c=1$ of Table~\ref{spin2table}: there we evaluated the anomalous dimension before implementing the normalization \eqref{normW2}, and the result is trivial because the 4-point correlation function is zero.

The anomalous dimensions in Table~\ref{spin2table} are sensitive to the spectrum of the seed CFT. In particular, these results are valid for symmetric orbifolds of the $\mathcal N = 2$ minimal models with $1 \le c < 3$, where the composition of the moduli on the cover is described in Sec.\,\ref{sec:mmmoduli}. To illustrate this sensitivity, let us focus on $n=3$. In this case, the moduli is given by \eqref{eq:modulin3}, which tell us that we are dealing with a descendant of a primary on the cover space; this is the structure used to get the appropriate entry in Table~\ref{spin2table}. If instead we considered $\tilde\phi_{(3)}$ to be a primary on the cover, the resulting (incorrect) anomalous dimension is
\be
\mu_{\textrm{incorrect}}=\frac{16 \pi^2 \lambda^2}{81}\frac{ (6 + c) (c N - 1)}{ c (c - 1)(N - 1)}~.
\ee
It is interesting to note that this value is always greater than or equal to $\mu_{(2)}$ for $n=3$, with the equality holding for $c=3$. 

As shown in Sec. \ref{sec:moduli}, symmetric orbifolds of CFTs with $3 \le c \le 6$ also admit 1/2-BPS marginal operators with twist $n = 2$. In these cases, we can obtain the anomalous dimension of $W_2$ from Table \ref{spin2table} provided that the seed CFT contains a primary field with dimension \eqref{eq:n2general} and charge $q = 1$. The anomalous dimension $\mu_{(2)}$ for symmetric orbifolds with $3 < c \le 6$ is then given by 
\eq{
\mu_{(2)}=\frac{3 \pi^2 \lambda^2}{64}\frac{ (24 + c)  (c N - 1)}{c (c - 1)(N - 1)}~.
}
In comparison with \cite{Gaberdiel:2015uca}, we find that if we take 
$c=6$, and $N\gg 1$, we obtain
\begin{align}
\mu_{(2)}\big|_{n =2,\, c = 6} =  \frac{9\pi^2\lambda^2}{32} + \mathcal O(1/N)\,. \label{n2c6spin2}
\end{align}
This result agrees with \cite{Gaberdiel:2015uca} up to a factor of 1/2.\footnote{The deformation parameter used in \cite{Gaberdiel:2015uca} is related to the one used here by $\lambda_{\textrm{there}} = N^{1/2} \lambda_{\textrm{here}}$.} 
Note that the same mismatch of 1/2 with respect to the results of \cite{Gaberdiel:2015uca} was found in \cite{Benjamin:2021zkn} for the lifting of certain spin-1 currents. 

We conclude this subsection by noting that the $\mathcal N = 2$ multiplet for the spin-2 current is denoted by $\mathcal{W}_{2}= \big\{W_2, \mathscr{W}_{5/2}^+, \mathscr{W}_{5/2}^-, \mathtt{w}_{3}\big\}$ whose individual components are
\eq{
\!\!\!\!\!\! W_2 &= T  - \frac{3}{2} (J J) + \frac{3(c N - 1)}{2c(N -1)}\sum_{j\ne l} J^{(j)}  J^{(l)}\, , \notag \\
\!\!\!\!\!\! \mathscr{W}_{5/2}^{\pm} & = \partial G^{\pm}  - 3 (J G^\pm)  + \frac{3(c N -1)}{c(N -1)} \sum_{j \ne l} J^{(j)} G^{\pm (l)}\,, \label{psiplus2} \\
\!\!\!\!\!\! \mathtt{w}_{3} & = \frac{3}{2}  \Big[ (G^+ G^-) - 2 (T J) - \partial T + \frac{2}{3} \partial^2 J \Big] -  \frac{3(c N -1)}{2c(N -1)}  \sum_{j \ne l} \Big[ G^{+(j)} G^{-(l)} - 2 T^{(j)} J^{(l)} \Big]\,.   \notag  
}
For convenience, we have not normalized these fields, a task that can be achieved by scaling each of them by a factor proportional to \eqref{normW2}. Since supersymmetry is protected under the deformation, the anomalous dimensions of $\mathscr{W}_{5/2}^{\pm}$ and $\mathtt{w}_{3}$ are simply given by the values in Table \ref{spin2table}.

\subsection{Spin-5/2 sector}\label{sec:fermionhs}

Generically, a symmetric orbifold contains one or more supercurrents $\mathcal W_s$ at each value of the spin $s$. One exception to this expectation is the spin-5/2 case. At spin-5/2, we can try to build the bottom components $W_{5/2}^{(\pm)}$ of candidate charged superfields $\mathcal W_{5/2}^{(\pm)}$ using the most general linear combination of terms with scaling dimension $5/2$, namely
\eq{
W_{5/2}^{(\pm)} = a_1  \partial G^{\pm}  + a_2 (JG^\pm)(z) + a_3 \sum_{j\ne l} J^{(j)}(z) G^{\pm (l)}(z)\,, \label{vp52}
}
where the $a_i$ coefficients are constants. It is not difficult to show that the only choice of coefficients that render these fields Virasoro-Kac-Moody primaries are given by the coefficients of $\mathscr{W}_{5/2}^{\pm}$ in \eqref{psiplus2}. Thus, the only holomorphic Virasoro-Kac-Moody primaries at spin-5/2 are the $\mathscr{W}_{5/2}^{\pm}$ currents of the spin-2 multiplet \eqref{psiplus2} and there are no spin-5/2 superprimary currents; their anomalous dimensions are simply given by Table \ref{spin2table}.

\subsection{Spin-3 sector}

Let us now consider the spin-3 sector. This is the first example where the fermionic currents $G^\pm$ contribute to the higher spin currents. At spin three, the symmetric orbifold Sym$^N({\cal C})$ has four Virasoro-Kac-Moody primaries but only two of these fields satisfy the conditions required for them to be the bottom component of a spin-3 superfield (see App.\,\ref{app:superprimary}). As a result, there are only two spin-3 superprimary multiplets that we denote by $\mathcal{W}_{3,1}$ and $\mathcal{W}_{3,2}$. The other two Virasoro-Kac-Moody primaries consist of a field built from the full $\mathcal N = 2$ currents of the symmetric orbifold --- i.e.~the spin-3 analog of $Y_2$ in \eqref{Y2def} --- and a descendant of the $\mathcal N = 2$ algebra acting on $W_2$. We have checked that the anomalous dimension of the former vanishes exactly, while that of the latter is given by the values in Table \ref{spin2table}, providing a consistency check of our calculations.

The bottom components $W_{3,a}$ of the spin-3 superprimary fields $\mathcal{W}_{3,a}$ can be conveniently written as
\eq{
W_{3,1}& = \sum_i \Big[ (J^{(i)}(J^{(i)} J^{(i)})) - 2 \, (T^{(i)} J^{(i)}) + \partial^2 J^{(i)} \Big]  \notag \\
& \hspace{15pt} + \frac{1}{c(N-1)} \sum_{i \ne j}\Big[ 3(2-c)\,J^{(i)} (J^{(j)} J^{(j)} ) + 3\, G^{+(i)} G^{-(j)} + 2c\, T^{(i)} J^{(j)} \Big] \notag \\
&  \hspace{15pt} + \frac{2(c-3)}{c(N-2)(N-1)} \sum_{i \ne j \ne k} J^{(i)} J^{(j)} J^{(k)} \,, \label{spin31}\\
W_{3,2} & = \sum_{i} \Big[ (G^{+(i)} G^{-(i)}) + \frac{4}{3} \, (T^{(i)} J^{(i)}) - \partial T^{(i)} - \partial^2 J^{(i)} \Big] \notag \\
&\hspace{15pt} - \frac{1}{6c(N - 1)} \sum_{i \ne j} \Big[ 24 \, J^{(i)} (J^{(j)} J^{(j)}) +  3 (13 + 2 c)\, G^{+(i)} G^{-(j)} + 4(9 + 2c) T^{(i)} J^{(j)}  \Big] \notag \\
& \hspace{15pt} + \frac{2(3 + 2c)}{c^2 (N - 2)(N - 1)} \sum_{i \ne j \ne k} J^{(i)} J^{(j)} J^{(k)} \,.
\label{spin32}
}
The first, second, and third lines in \eqref{spin31} and \eqref{spin32} correspond to linear combinations of single-trace, double-trace, and triple-trace operators. In particular, note that double and triple-trace operators are suppressed by factors of $1/N$ and $1/N^2$ respectively, such that their contributions are negligible in the large-$N$ limit. It is also interesting to note that as we increase the spin $s$, the higher spin currents of the symmetric orbifold feature higher order multi-trace operators that consist not only of the $\mathcal N = 2$ currents, but also of higher spin $\mathcal W_{s'}$ currents with $s' < s$. 

We have written the spin-3 currents \eqref{spin31} and \eqref{spin32} in a basis that favours the single-trace contributions of $J$ and $G^\pm$, respectively, but is not orthogonal. An orthogonal basis can be obtained from the following linear combination of currents
\eq{
\widetilde W_{3,1} & \coloneqq (13 + 2c) W_{3,1} + 6 W_{3,2}\,, \qquad \widetilde W_{3,2} & \coloneqq 4(3 + c N) W_{3,1} + c N (6 - c N) W_{3,2}\,. 
}
In this basis, the norms of the spin-3 currents are given by
\eq{
\!\!\! \langle \widetilde W_{3,1} | \widetilde W_{3, 1}\rangle &= \frac{2N(c-1)(c+6)(2c -3)\big[72 - 6 c N(9 + 2c)  + c^2 N^2(13 + 2c)\big] }{9c (N - 2)(N - 1)}\,, \label{spin31norm} \\
\!\!\! \langle \widetilde W_{3,2} | \widetilde W_{3, 2}\rangle &= \frac{N(cN\!+\!6)(2cN \!-\!3)(5cN \!-\!12)\big[72 - 6 c N(9 + 2c)  + c^2 N^2(13 + 2c)\big] }{27(N - 1)}\,, \label{spin32norm}
}
while $ \langle \widetilde W_{3,1} | \widetilde W_{3, 2}\rangle$ vanishes by construction. We see that the $\widetilde W_{3,1}$ current becomes null whenever $c = 1$ or $c = 3/2$. In these cases, that include two of the $\mathcal N = 2$ minimal models, this current must be modded out from the spectrum. Consequently, it is not going to be surprising  to find below that for the $c =1$ and $c = 3/2$ minimal models, one of the spin-3 anomalous dimensions vanishes exactly (before normalization). As discussed in Sec.~\ref{se:spin2sector}, this follows from the fact that correlation functions of a null state with any other operator must vanish. 

Using the methods described in Sec.~\ref{sec:coverbasemap} we obtain the matrix $\gamma_{(3)}$ of spin-3 anomalous dimensions for each of the twisted moduli of the symmetric orbifolds of the $\mathcal N = 2$ minimal models. The eigenvalues of this matrix are denoted by $\mu_{(3)}^{(\pm)}$ and satisfy
\eq{
\textrm{det}\, \gamma_{(3)} =  \mu_{(3)}^{(+)}\mu_{(3)}^{(-)}, \qquad \textrm{tr}\, \gamma_{(3)} =  \mu_{(3)}^{(+)} + \mu_{(3)}^{(-)}.
}
For values of the seed central charge in the range $1 \le c \le 2$, where the symmetric product orbifold admits twisted moduli with values ranging up to $n=7$ (see Table \ref{t:moduli}) the eigenvalues $\mu_{(3)}^{(\pm)}$ are reported in Table~\ref{spin3table}. On the other hand, when the seed central charge lies in the range $2 < c < 3$, we have only twist $n = 2,3$ moduli, and the eigenvalues $\mu_{(3)}^{(\pm)}$ can be obtained from the explicit expressions for the  $\gamma_{(3)}$ matrices given below.
\begin{description}[leftmargin=0cm]
\item[Twist-2.] Using the fact that the composition of the moduli is of the form \eqref{eq:n2general},  the determinant and trace of $\gamma_{(3)}$ read
\eqsp{
\textrm{det}\, \gamma_{(3)} &= \frac{27 \pi ^4  \lambda ^4}{16384 }\frac{ (c N + 6) (2 c N - 3) (-4320 +2412 c + 276 c^2 + 7 c^3)}{c^2 (c-1)  (c+6) (2 c-3) (N-1)^2} \,,  \\
\textrm{tr}\,\gamma_{(3)} &= \frac{3 \pi ^2 \lambda ^2}{128}\frac{1}{ c (c-1) (c+6) (2 c-3) (N-1)} \Big( 4212 - 4284 c + 891 c^2 + 71 c^3  \\ 
&\qquad\qquad+ cN  (-1044 + 72 c + 377 c^2 + 20 c^3) \Big) \,. \label{n2spin3}
}
For minimal models this covers theories with $c = 3k/(k+2)$ with $k = 4, 6, \dots$. This result is also valid for theories with $3\leq c\leq 6$ for which a suitable primary exists that conforms with \eqref{eq:n2general}. 

Reading off $\mu_{(3)}^{(\pm)}$ from \eqref{n2spin3} is straightforward, but the expressions are cumbersome to write down. Instead, we report their values at large $N$, which are given by
\eq{
\!\!\!\!\!\!\! \mu_{(3)}^{(\pm)}& = \frac{3\pi^2\lambda^2}{256 (c-1)(c+6)(2c -3)} \Big(\!-\!1044 + 72 c + 377 c^2 + 20 c^3 \notag\\ 
& \hspace{15pt} \pm \sqrt{2956176 - 3991680 c + 1387512 c^2 - 9504 c^3 - 12599 c^4 + 656 c^5 +  64 c^6}\, \Big) \notag \\
& \hspace{15pt}+ {\cal O}(1/N)~.
}
It is worth mentioning that this expression is always positive for the range $2\leq c\leq 6$.
\item[Twist-3.] Using that the composition of the moduli is of the form \eqref{eq:n3general},  the determinant and trace of $\gamma_{(3)}$ are now given by
\eq{
\textrm{det}\, \gamma_{(3)} &= \frac{8 \pi^4 \lambda^4}{2187}\frac{(c N + 6) (2 c N - 3)}{ c^5 (c - 1) (c + 6) ( N -2) (N - 1)^2} \Big( -324 + 3159 c^2 - 3603 c^3 \notag \\
  & \hspace{15pt} - 423 c^4 - 9 c^5   - N (108 c + 1134 c^2 - 1809 c^3 - 158 c^4 - 3 c^5) \Big)   \,,  \notag \\
\textrm{tr}\,\gamma_{(3)} &=  \frac{2\pi^2 \lambda^2 }{81}\frac{1}{c^2 (c-1)  ( c +6) ( N-2) ( N-1)} \Big( -6 (234 - 693 c + 316 c^2 + 7 c^3) \notag\\
& \hspace{15pt} + N (378 - 873 c + 282 c^2 - 701 c^3 - 30 c^4)  \label{n3spin3} \\
& \hspace{15pt} +  2 c N^2 (-234 + 189 c + 173 c^2 + 6 c^3)  \Big) \,. \notag
}
These expressions apply for minimal models with $c = 3k/(k+2)$ with $k \ge 3$. They are also valid for theories with $c=3$, the maximal allowed value for a twist-3 deformation of an ${\cal N}=2$ CFT$_2$. At large $N$, the eigenvalues of \eqref{n3spin3} are given by
\begin{align}
    \begin{split}
\mu_{(3)}^{(\pm)}& = \frac{2\pi^2\lambda^2}{81c(c-1)(c+6)}  \Big( -234 + 189 c + 173 c^2 + 6 c^3  \\
&\hspace{140pt}\pm (5c-3) \sqrt{5220 - 780 c + 49 c^2} \, \Big)    
 + {\cal O}(1/N)\,,
\end{split}
\end{align}
which are positive for all $c$ in the range $9/5\le c\le3$.
\end{description}

  \begin{table}[ht!]
  \begin{center}\def\arraystretch{1.65}
    \addtolength{\leftskip} {-2cm}
    \addtolength{\rightskip}{-2cm}
  \begin{tabular}{|c|c|c|p{3.80cm}<{\centering}|c|}
\hline
  $k$ & $c$ & $n$ & $\mu_{(3)}^{(+)}$ & $\mu_{(3)}^{(-)}$ \\ 
  \hline 
  \multirow{2}{*}{1} &  \multirow{2}{*}{1} & 5&   \multirow{2}{*}{---}  & $ \frac{2322  \pi ^2 \lambda ^2  (6 + N) (-3 + 2 N) }{3125 (N -2)(N -1)}$    \\
  \cline{3-3}\cline{5-5}
  &  & 7 & &$ \frac{2466\pi^2 \lambda^2 (6 + N) (-3 + 2 N)}{2401(N-2)(N-1)}$   \\
 \hline
   \multirow{3}{*}{2} &  \multirow{3}{*}{$\frac{3}{2}$} & 3&   \multirow{3}{*}{---} &  $ \frac{4\pi^2 \lambda^2 (4 + N)( - 9 + 5 N)}{27 (N-2)(N-1)}$ \\
  \cline{3-3}\cline{5-5}
  &  & 4 & &$\frac{\pi ^2 \lambda ^2 (4 + N) (-1876 + 1079 N)}{1024 (N -2)(N -1)}$  \\
    \cline{3-3}\cline{5-5}
  &  & 5 &  &$\frac{4\pi ^2 \lambda ^2 (4 + N) (-1789 + 1040 N)}{3125 (N -2)(N -1)}$   \\
  \hline
 3 & $\frac{9}{5}$ & 3 & \multicolumn{2}{c|}{$\frac{\pi^2 \lambda^2 (-1780 -27100 N + 14619 N^2 \pm \sqrt{1579226000 - 2952279840 N + 2003396824 N^2 - 584464200 N^3 + 62105625 N^4} )}{9477(N-2)(N-1)}$}  \\
 \hline
   \multirow{3}{*}{4} &  \multirow{3}{*}{2} & 2 & \multicolumn{2}{c|}{$\frac{3 \pi ^2 \lambda ^2 (- 7 + 48 N   \pm \sqrt{2857 -3480 N + 1056 N^2}\,)}{128 (N-1)}$}  \\
  \cline{3-5}
  &  & 3 &  \multicolumn{2}{c|}{   $\frac{\pi^2 \lambda^2 (-126 - 791 N + 442 N^2 \pm \sqrt{1288548 - 2376252 N + 1584865 N^2 - 453460 N^3 + 47236 N^4} )  }{324(N-2)(N-1)}  $    } \\
  \cline{3-5}
   &  & 4 &  \multicolumn{2}{c|}{$\frac{3\pi^2 \lambda^2 (  -318 - 879 N + 549 N^2 \pm 
 \sqrt{3} \sqrt{
  754068 - 1275900 N + 753741 N^2 - 182840 N^3 + 15643 N^4})}{1024 (N - 2) (N - 1)} $} \\
   \hline
  \end{tabular}
  \caption{The anomalous dimensions of the spin-3 currents of symmetric orbifolds with seed central charge $1 \le c \le 2$. When $c = 1$ or $3/2$ one of the spin-3 currents becomes null and the anomalous dimension of the unnormalized current vanishes. The non-vanishing values of the anomalous dimensions are real and positive for all $N \ge n$.}
  \label{spin3table}
  \end{center}
  \end{table}
  
It is interesting to compare the anomalous dimensions of the spin-2 and spin-3 currents. For convenience we work in the large-$N$ limit. Since the $W_2$ current becomes null when the seed central charge is $1$, we only consider the cases where $1 < c < 3$. For each of the twisted moduli of the symmetric orbifold we find that
\begin{itemize}
\item when $c = 3/2$ and there is only one spin-3 current, the anomalous dimensions satisfy
\eqst{
\mu_{(2)} > \mu_{(3)}^{(-)}\,, \qquad (N \gg 1)~;
}
\item while for $3/2 < c < 3$ we instead have
\eqst{
 \mu_{(3)}^{(+)} > \mu_{(2)} > \mu_{(3)}^{(-)}\,, \qquad (N \gg 1)~.
}
\end{itemize}
Once the higher spin currents become massive we expect their anomalous dimensions to fall into Regge trajectories such that $\mu_{(s+1)} > \mu_{(s)}$ for each spin $s$. The results above suggest that this is the case for the spin-2 current and the spin-3 current associated with $\mu_{(3)}^{(+)}$, while the current associated with $\mu_{(3)}^{(-)}$ may be interpreted as lying in a different Regge trajectory. In order to verify if this is indeed the case it will be necessary to compute the anomalous dimensions of additional higher spin currents.
  
In analogy with the spin-2 case considered in Sec.~\ref{se:spin2sector}, the anomalous dimensions of the spin-3 currents can be extended to symmetric orbifolds with seed central charge $3 \le c \le 6$. As described earlier, this is possible because these theories admit 1/2-BPS marginal operators at twist $n = 2$. Assuming that these theories contain a primary field with weight \eqref{eq:n2general} and charge $q = 1$, the anomalous dimensions of symmetric orbifolds with $3 \le c \le 6$ can be obtained from the eigenvalues of $\gamma_{(3)}$ given in \eqref{n2spin3}. In particular, let us consider the $c = 6$ case, where the eigenvalues $\mu_{(3)}^{(\pm)}$ are simply given by
\eq{
\mu_{(3)}^{(+)} \big|_{n = 2, c = 6} & = \frac{3\pi^2 \lambda^2}{64}\frac{ ( 2 + 8 N + \sqrt{19 - 13 N + 4 N^2})}{ (N -1)} = \frac{15\pi^2 \lambda^2}{32} + \mathcal O(1/N)\,,  \label{n2c6spin31}\\
\mu_{(3)}^{(-)} \big|_{n = 2, c = 6} &= \frac{3 \pi^2 \lambda^2}{64}\frac{(2 + 8 N - \sqrt{19 - 13 N + 4 N^2})}{(N -1)} = \frac{9\pi^2 \lambda^2}{32}  + \mathcal O(1/N)\,.  \label{n2c6spin32}
}
It is interesting to note that in the large-$N$ limit, $\mu_{(3)}^{(-)}$ agrees with the anomalous dimension of the spin-2 current $\mu_{(2)}$ in \eqref{n2c6spin2}. More generally, we find that for $c = 6$ the anomalous dimensions satisfy $\mu_{(3)}^{(+)} > \mu_{(3)}^{(-)} \ge \mu_{(2)}$ where the last inequality is saturated when $N \to \infty$.

We conclude by noting that the spin-3 currents \eqref{spin31} and \eqref{spin32} are the bottom components of two $\mathcal N = 2$ superprimary fields, each of which consist of two additional charged spin-7/2 fields and a neutral spin-4 field. Since supersymmetry is preserved by the deformation, the anomalous dimensions of all of these fields --- or more precisely, of the linear combinations that diagonalize the matrix of anomalous dimensions --- are also given by Table \ref{spin3table} and the eigenvalues of~\eqref{n2spin3} and \eqref{n3spin3}.

\section*{Acknowledgements}

We thank Nathan Benjamin and Jildou Hollander for interesting discussions, and collaborations on related topics.
The work of LA, SB and AC is supported by the Delta ITP consortium, a program of the Netherlands Organisation for Scientific Research (NWO) that is funded by the Dutch Ministry of Education, Culture and Science (OCW). The work of LA was also supported by the European Research Council under the European Union's Seventh Framework Programme (FP7/2007-2013), ERC Grant agreement ADG 834878. The work of AB is supported by the NCCR 51NF40-141869 The Mathematics
of Physics (SwissMAP). The work of CAK is supported in part by the Simons Foundation Grant No. 629215 and by NSF Grant 2111748. AB and SB acknowledge the workshop ``Qubits on the Horizon 2" where fruitful discussions occured.

\appendix

\section{Conventions}\label{app:conventions}
In this appendix we establish our conventions for the $\mathcal N = 2$ superconformal algebra, introduce its superprimary fields, and describe the spectrum of the $\mathcal N = 2$ minimal models.

\subsection{\texorpdfstring{${\cal N}=2$}{N=2} superconformal algebra}\label{app:N2SCA}
Let us begin by describing basic properties of the $\cN=2$ superconformal algebra. This algebra is characterized by the stress tensor $T$,  a weight-1 $U(1)$ current $J$, and two weight-3/2 fermionic currents $G^+$ and $G^{-}$ with $U(1)$ charges $+1$ and $-1$, respectively. In terms of its generators, the corresponding commutators are given by
\begin{equation}
\begin{aligned}\label{superVir comms}
\bigl[L_m,L_n\bigr]
 ={}&
  (m-n)L_{m+n} +\frac{c}{12}m(m^2-1)\delta_{m+n,0}\,,
   \\
\bigl[J_m,J_n\bigr]
={}&
\frac{c}{3}m\,\delta_{m+n,0}\,,
  \\
\bigl\{G^+_r,G^-_s\bigr\}
 ={}&
2L_{r+s}+(r-s)J_{r+s} +\frac{c}{3}\Big(r^2-\frac{1}{4}\Big)\delta_{r+s,0}\,,
 \\
\bigl[L_m,J_n\bigr]
={}&
 -n\, J_{m+n}\,, \\ 
 \bigl[J_m,G^\pm_r\bigr]
  ={}&
  \pm G^\pm_{m+r}\,,
 \\
  \bigl[L_m,G^\pm_r\bigr]
  ={}&
   \Big(\frac{m}{2}-r\Big) G^\pm_{m+r}\,,
   \\
\bigl\{G^\pm_r,G^\pm_s\bigr\}
={}&
0\ .
\end{aligned}
\end{equation}
The standard Hermiticity properties of the generators on the plane are \cite{Blumenhagen:2009zz,Blumenhagen:2013fgp}
\begin{equation}\label{super Virasoro Hermiticity}
\begin{aligned}
\left(L_{n}\right)^{\dagger} = L_{-n}\,,\qquad \left(J_{n}\right)^{\dagger} = J_{-n}\,,\qquad \left(G^{+}_{r}\right)^{\dagger} = G^{-}_{-r}\,.
\end{aligned}
\end{equation}
 In the Ramond (R) sector the fermionic generators $G^{\pm}_{r}$ are integer-modded, $r \in \Z$, while in the Neveu-Schwarz (NS) sector they are half-integer-modded, $r \in \Z + \frac{1}{2}$.

The $\cN=2$ superconformal algebra is invariant under the so-called spectral flow automorphism \cite{Schwimmer:1986mf}, which reads
\begin{equation}
\begin{aligned}\label{spectral flow 1}
L_n &\quad \to \quad\,\,\,\, L'_{n} ={} L_n + \eta J_{n} + \frac{\eta^{2}}{6}c \delta_{n,0}\,,
\\
J_n &\quad \to  \quad \,\,\,\, J'_{n}  ={} J_n + \frac{c}{3}\eta \delta_{n,0}\,,
\\
G^{\pm}_{r} &\quad \to \quad  G_{r}^{\pm '}  ={} G^{\pm}_{r\pm \eta}\,,
\end{aligned}
\end{equation}
where $\eta$ is a continuous parameter. For $\eta \in \mathbb{Z} + 1/2$ the flow interpolates between the NS and R sectors, while for $\eta \in \mathbb{Z}$ it maps the NS and R sectors to themselves.

The zero modes $L_0$ and $J_0$ commute such that primary states are labeled by the eigenvalues $h$ and $q$ of these operators, namely,\footnote{We recall that for ${\cal N}=4$ algebras $q\in \Z$, while for ${\cal N}=2$ the charge $q$ is a rational number.} 
\begin{equation}\label{superVirasoro Cartan}
L_0|\vp\rangle=h|\vp\rangle\,,\qquad J_0|\vp\rangle=q|\vp\rangle\,.
\end{equation}
Furthermore, primary states satisfy the usual highest-weight conditions
\begin{equation}
\begin{aligned}\label{superVirasoro hw conditions 1}
 G^{\pm}_{r}\bigl |\vp\bigr\rangle ={}&
  0\, ,&
  &\qquad\quad &
   r >{}&0\,,
     \\
  L_{n}\bigl |\vp\bigr\rangle ={}& J_{n}\bigl |\vp\bigr\rangle = 0\,,&
    &\qquad\quad &
   n >{}&0\,.
\end{aligned}
\end{equation}

In the NS sector of the Hilbert space of an $\mathcal{N}=2$ SCFT, it is useful to define chiral and anti-chiral states to be those which, in addition to \eqref{superVirasoro hw conditions 1}, respectively satisfy
\begin{align}\label{eq:chiral}
	G^{+}_{-\frac{1}{2}}\left|\vp\right\rangle&=0\,,
\end{align}
and 
\begin{align}\label{eq:antichiral}
	G^{-}_{-\frac{1}{2}}\left|\vp\right\rangle&=0\,.
\end{align}
Using the mode algebra one can easily prove (see e.g. \cite{Blumenhagen:2009zz}) that $\left|\vp\right\rangle$ is an $\mathcal{N}=2$ \mbox{(anti-)}chiral primary if and only if $h = q/2$ ($h=-q/2$). As usual, chiral representations correspond to short supermultiplets.

Finally, it is useful to record the unitarity bounds of the ${\cal N}=2$ algebra \cite{Boucher:1986bh}. For states in the NS sector unitarity restricts $h$ to satisfy
\begin{align}\label{eq:NSbound}
  h \geq r q  + \frac{(c-3)}{24}\left(1-4r^{2}\right)\,,\qquad r \in \mathbb{Z} +\frac{1}{2}\,,  
\end{align}
while for the R-sector we have
\begin{align}\label{eq:Rbound}
h \geq nq  + \frac{c}{24}\left(1-4n^{2}\right) + \frac{n(n-1)}{2}\,,\qquad n \in \mathbb{Z}\,.
\end{align}
These sets of linear relations are obtained from the unitary conditions that follow from the Kac determinant. The (half-)integer spacing comes  from incorporating spectral flow sectors in the bounds. 

\subsection{Superprimary fields}\label{app:superprimary}

In this appendix we describe the superprimary fields of the $\mathcal N = 2$ superconformal algebra and write down their OPEs with the conserved currents following the conventions of \cite{Melnikov:2019tpl}. We focus on the conditions for the holomorphic sector but there is of course a natural counterpart for the anti-holomorphic sector.  

We denote a generic superprimary field as $\Phi_{(h, q)}$, which is characterized by its conformal weight $h$ and $U(1)$ charge $q$. This superfield  consists of four components $\Phi_{(h, q)} = \{ \vp , \psi^+, \psi^-, \chi\}$ which in the NS sector satisfy
\begin{equation}\label{eq:def-superprim}
\begin{aligned}
L_0 \Ket{\vp} &= h \Ket{\vp}\,, \qquad\qquad\, J_0 \Ket{\vp} = q \Ket{\vp}\,, \\ 
L_n \Ket{\vp} &= G^\pm_{-1/2 +n} \Ket{\vp} = 0 \,,\qquad  \textrm{for} \,\, n > 0\,, \\
\Ket{\psi^{\pm}} &= \mp G^{\pm}_{-1/2} \Ket{\vp}\,, \\ 
\Ket{\chi} &= G^+_{-1/2} G^-_{-1/2} \Ket{\vp} - L_{-1} \Ket{\vp}~.
\end{aligned}
\end{equation}
It is clear from these expressions that the states $\vp$ and $\psi^\pm$ are Virasoro-Kac-Moody primaries, while $\chi$ is a Virasoro-Kac-Moody descendant. It follows that the OPEs between the currents in the ${\cal N}=2$ algebra and the components of the superprimary field are
\begin{equation}
    \begin{aligned}
T(z) \vp(0) &\sim \frac{h \vp(0)}{z^2} + \frac{\p \vp(0)}{z}\,, \\ 
T(z) \psi^{\pm}(0) &\sim \frac{(h + \tfrac{1}{2}) \psi^{\pm}(0)}{z^2} + \frac{\p\psi^{\pm}(0)}{z}\,, \\
T(z) \chi(0) &\sim \frac{q\vp(0)}{z^3} + \frac{(h+1) \chi(0)}{z^2} + \frac{\p \chi(0)}{z}\,, \\
J(z) \vp(0) &\sim \frac{ q \vp(0)}{z}\,, \\
J(z) \psi^{\pm}(0) &\sim \frac{(q \pm 1)\psi^{\pm}(0)}{z}\,, \\
J(z) \chi(0) &\sim \frac{2 h \vp(0)}{z^2} + \frac{q \chi(0)}{z}\,, \\
 G^\pm(z) \vp(0) & \sim \mp \frac{\psi^{\pm}(0)}{z}\,, \\
G^{\pm}(z) \psi^{\pm}(0) &= 0\,, \\
G^{\pm}(z) \psi^{\mp}(0) & \sim \pm\frac{(2h \pm q) \vp(0)}{z^2} + \frac{\chi(0) \pm \p \vp(0)}{z}\,, \\
G^{\pm}(z) \chi(0) &\sim \frac{(2h + 1 \pm q) \psi^{\pm}(0)}{z^2} + \frac{\p \psi^{\pm} (0)}{z}\,. \label{OPEs}    
    \end{aligned}
\end{equation}

If we implement the conditions \eqref{eq:chiral} and \eqref{eq:antichiral}, there is a shortening of the multiplet: this occurs when $\varphi$ is a (anti-)chiral primary, as defined in \eqref{eq:chiral} and \eqref{eq:antichiral}. In the chiral case we impose $\psi^+ =0$ while in the anti-chiral one we impose  $\psi^- = 0$ instead. In both situations $\chi$ becomes a descendant of $\vp$. From \eqref{eq:def-superprim}, the fields making up the short supermultiplet are identified as
\eq{
\Phi_{(h, 2h)} = \{\vp , 0, \psi^-, \p \vp \}\,, \qquad \Phi_{(h, -2h)} = \{\vp ,  \psi^+, 0, -\p \vp \}\,,
}
which correspond to a chiral and an anti-chiral primary, respectively. 

\subsection{Spectrum of  \texorpdfstring{${\cal N}=2$}{N=2}  minimal models}\label{app:minmodel}

For theories with $1\leq c<3$ there is a complete classification of SCFTs with ${\cal N}=(2,2)$ supersymmetry --- these are the  ${\cal N}=2$ minimal models. The classification is possible because there is a finite number of irreducible representation for a fixed value of $c$ in this range  \cite{Boucher:1986bh, DiVecchia:1986fwg}. Here we will list the values of the $U(1)$ charge and the dimension of the highest weight state for each representation. This information is used in the main text to identify the marginal operator $\Phi$ on the cover space. 

The $U(1)$ charges and the dimension of the highest weight states in the Ramond sector are given by 
\be\label{QR}
h_{r,s}^\epsilon=\frac{r^2-s^2}{4(k+2)}+\frac{c}{24}~ ,\qquad  q_s^\epsilon=\frac{s}{k+2}+\frac{\epsilon}{2}~.
\ee
The labels run as
\be
r\in \{1, \ldots  k+1\}~,\qquad \ 0\leq |s+\epsilon|\leq r-1~, \qquad
\ r+ s \equiv 0 \pmod 2\ ,
\ee
where $\epsilon=\pm$ and the positive integer $k$ labels the central charge of the SCFT via \eqref{eq:cmm}.  From (\ref{QR}) we see that the BPS representations, i.e.~the Ramond ground states, are given by $r=|s|$. The highest weight states in the NS sector can be easily obtained using the spectral flow relation \eqref{spectral flow 1} on the zero modes of the algebra. 

\section{\texorpdfstring{$\cN=2$}{N=2} Ward identities and correlation functions}\label{app:WI}
In this appendix we explain how we use Ward identities to compute correlation functions of the form \be
\langle \phi_1 | X^1(z_1) X^2(z_2)\cdots X^n(z_n)|\phi_2\rangle\ ,
\ee
where the $X^i(z)$ are  chiral fields of the $\cN=2$ superconformal algebra and the $\phi$ are general descendant fields. 

Our general strategy is to split the fields $W(z)$ into a creator and an annihilator part,
\be
X(z) = X^c(z)+X^a(z) = \sum_{r+h<1} X_r z^{-r-h}+\sum_{r+h\geq 1}X_r z^{-r-h}\, .
\ee
We then (anti-)commute the annihilator part all the way to the right, and the creator part all the way to the left. In doing so we pick up (anti-)commutators $[X^{c,a}(z_i),X(z_j)]_\pm$.
These commutators are very similar to the Ward identities coming from the OPE of $X(z_i)$ with $X(z_j)$. They do however also contain all non-singular terms.
Choosing the cutoff between annihilator and creator terms is somewhat arbitrary, but we made our choice to give the nicest form to the commutators.
Finally, the creator and annihilator parts act on the $\phi$ states as
\be
X^a(z)|\phi_2\rangle = \sum_{r\geq-h_X+1}^{N_2}z^{-r-h_X} |X_r \phi_2\rangle\,,
\ee
and 
\be
\langle \phi_1| X(z) = \sum^{N_1}_{r>h_X-1} z^{r-h_X}\langle X^\dagger_r \phi_1 |\, ,
\ee
where $N_{1,2}$ is the total weight added by the descendants in $\phi_{1,2}$.

Let us now discuss how this works in practice. In the NS sector, the (anti-)commutators turn out to precisely match the singular part of the Ward identity OPEs. This is not surprising: In that case, the correlation function is a meromorphic function in the $w_i$. Subtracting all poles in the OPEs thus leaves a bounded function with no poles, which by Liouville's theorem has to be constant. The constant in turn has to be zero because of cluster decomposition. In summary this means that the correlator is indeed what is expected by pole subtraction.

The situation in the Ramond sector is slightly more complicated. The correlator is no longer a meromorphic function: the fermionic variables have branch cuts from 0 to $\infty$. The pole subtraction argument used above thus no longer works. This shows up in the anti-commutator of $G^+$ with $G^-$.
In this case we have
\begin{align}
\begin{split}
\{G^{a+}& (z), G^-(w) \} = (zw)^{-3/2}\sum_{r\geq 0,s\in\mathbb{Z}} \{G^+_r,G^-_{-s}\} z^{-r} w^{s}  \\
& = (zw)^{-3/2}
\sum_{r\geq 0,n\in\mathbb{Z}} \left(2 L_n + (2r-n)J_n +\frac{c}{3} \left(r^2-\frac{1}{4}\right) \delta _{0,n}\right)w^{-n}(w/z)^r  \\
& = (w/z)^{1/2}\left(\frac{2T(w)}{z-w} + \frac{2J(w)}{(z-w)^2} + \frac {\partial(wJ(w))}{w(z-w)}
+ \frac c{12}\frac{3w^2+6wz-z^2}{(z-w)^3w^2}
\right), 
\end{split}
\end{align}
where in the second line we substituted $s=r-n$.
A similar computation shows that
\be
\{G^{c+}(z),G^-(w)\}= - \{G^{a+}(z), G^-(w) \}\ ,
\ee
as is needed for consistency.
Note that the factor of $(w/z)^{1/2}$ is exactly what is needed to get the correct branch cuts at $0$ and $\infty$ for $w$ and $z$.

It may seem surprising that the Ramond sector Ward identities look so much more complicated than the NS ones. However, this is only due to the fact that our Ward identities are exact, and not just up to singular terms. In fact, if we are only interested in the singular terms as $z\to w$, we recover the usual Ward identity:
\be
\{G^{a+}(z), G^-(w) \} = \frac{2T(w)}{ (z-w)}+\frac{2J(w)}{(z-w)^2}+\frac{\partial J(w)}{ (z-w)}+
\frac{2c}{3 (z-w)^3}+O\left((z-w)^0\right).
\ee
Similar regular terms appear if we collide $G(z)$ with a bosonic operator at $w$, \ie if we do the expansion $z\to w$. However, we can avoid this by colliding the bosonic operator with $G(z)$ instead, \ie by taking the expansion $w\to z$ instead. By the above argument this expansion will only contain singular terms.

\section{Evaluating symmetrized correlation functions}\label{app:largeN}
In this appendix we provide additional details on the evaluation of correlation functions in symmetric product orbifolds.

\subsection{\texorpdfstring{$N$}{N} dependence of correlation functions}
We follow Sec.\,4 of \cite{Benjamin:2021zkn}:
operators in the symmetric orbifold are given by orbits under the action of $S_N$.
We will denote by $\check O$ a representative of that orbit, usually chosen such that the all non-vacuum factors are up front. The actual symmetrized operator $O$,  up to normalization, is then given by 
\be\label{prestateapp}
O = \sum_{g\in S_N} g\check O\ .
\ee
More precisely, pick a representative of this orbit $\check O$ whose first $l$ factors are non-vacuum factors, and whose last $N-l$ factors are vacua,
\be
\check O = O^1\otimes \cdots O^l\otimes \bigotimes^{N-l}\vac\ .
\ee
Note we can use $\check O$ to construct such a state for any $N$, as long as $N\geq l$. This allows to construct a large-$N$ limit.
A normalized state which is permutation symmetric can then be written as the normalized orbit of $\check \phi$ with
\be
||O||\sim \sqrt{N! (N-l)!}
\ee
If $l=1$, then $\phi$ is a single-trace state, if $l=2$, double trace \etc
If $\phi$ is a linear combination of states of different length, then define $L=\min(l)$. We will call the term of length $L$ the \emph{head} of the state, and terms with $l>L$ the \emph{tail}. We will also call $L$ the \emph{head length}. If the coefficients in front of the different multi-trace components written as in (\ref{prestateapp}) are independent of $N$, then $L$ then fixes the normalization to leading order as $\sqrt{N! (N-L)!}$. 

Note that this argument only works as long as the coefficients in the linear combination do not depend on $N$, which can fail in some common cases. Take for instance the field $JJ$. Written as in (\ref{prestateapp}) we have 
\be
(\sum_i J^i)(\sum_j J^j) = ((N-1)!)^{-1}\sum_{g\in S_N}g (J^1 J^1+(N-1)J^1J^2)\ .
\ee
We see that the double trace term comes with a factor $N-1$, and will therefore also contribute to the overall normalization.

For simplicity let us assume that $\phi$ only has terms of length $l=L$. Somewhat symbolically, the 4pt function we need can be computed as
\be\label{Iexpression}
I(x,\xb) = \sum_{\rho_1,\rho_2} \kappa_{\rho_1}(N) \sum_{g\in S_{K_\rho}}\langle G_{-1/2} \varphi|\check O^{(1,\rho_1)}(1,1)g \check O^{(2,\rho_2)}(x,\xb)|G_{-1/2} \varphi\rangle~.
\ee
Here $\rho_1$ is a configuration $\rho_1=(a_1,a_2,\ldots, a_n,S)$ that consists of an ordered tuple of factors $(a_1,a_2,\ldots, a_n)$ describing which factors of $\check O^1$ sit in the first copies (1) through ($n$), and an unordered multiset $S$ which simply contains all remaining $N-n$ factors of $O^i$. $K_{\rho_1}$ is the number of non-vacuum factors in $S$, such that $K_{\rho_1}$ is between $L_1$ and $L_1-n$. Similarly $\rho_2$ is a configuration that describes $O^2$. Since the correlator vanishes unless the same number of vacuum factors appear in the last $N-n$ copies, we write $K_\rho = K_{\rho_1}=K_{\rho_2}$.

The crucial point of (\ref{Iexpression}) is that all sums that appear are independent of $N$. The only $N$-dependence comes from $\kappa_\rho(N)$. This factor was worked out in \cite{Benjamin:2021zkn} for $n=2$. Their argument can be easily generalized to give
\be\label{Ndepfactor} 
\kappa_\rho(N) = 
\frac{(N-n)!(N-l_1)!(N-l_2)!}{||O^1|| ||O^2||(N-K_\rho-n)!}~.
\ee

To discuss the large-$N$ behavior of states $O$ with mixed trace length, let us assume that the normalization is fixed by the head length, $||O||\sim \sqrt{N! (N-L)!}$. For the contribution of two pieces of trace length $l$ we then get
\be
\kappa_\rho(N) 
\sim \frac{(N-n)!(N-l_1)!(N-l_2)!}{N!\sqrt{(N-L_1)!(N-L_2)!}(N-K_\rho-n)!}
\sim N^{-l_1-l_2+\frac12L_1+\frac12L_2+K}
\ee
Obviously for fixed states, the leading contribution comes from picking $K$ as large as possible. However, if $K=l_1$ (or $l_2$), then the correlation function is actually disconnected, since $O^{1}$ is orthogonal to the twist fields which live in $(1\cdots n)$. The maximal connected contribution is thus from $K$ given by
\be
K = \min(l_1,l_2)-1\ .
\ee
Using 
\be
l_1+l_2-2\min(l_1,l_2)= |l_1-l_2|\ ,
\ee
we get that the contribution is
\be
N^{\frac12(L_1-l_1)+\frac12(L_2-l_2)-\frac12|l_1-l_2|-1}\ .
\ee
From this we immediately find the following results:
\begin{itemize}
	\item If $L_1=L_2$, then the leading contribution comes from the two head terms, and is $\mathcal{O}(N^{-1})$. Any tail terms will give subleading contributions.
	\item If $L_1\neq L_2$, then the contribution is
	\be
	\mathcal{O}(N^{-\frac12|L_1-L_2|-1})\ ,
	\ee
	that is subleading in $1/N$. This contribution however can potentially also come from tail terms. 
\end{itemize}
In summary, we can say the following about the matrix elements between states of mixed trace length: 
\begin{itemize}
	\item To leading order it is enough to keep the head terms of a state. 
	\item Matrix elements between states of different head length are always subleading.
	\item However, genuine multi-trace states with no single trace terms can still give leading contributions amongst themselves. Their matrix elements with single-trace states however are subleading.
\end{itemize}

\subsection{Evaluating single-trace sums}\label{sec:sums}
Let us now describe how to evaluate the sums in (\ref{Iexpression}) in practice, for two single-trace terms of the form
\be\label{eq:basecfst}
\langle G_{-1/2} \varphi|O^{(j)}(1)O^{(j')}(x)|G_{-1/2} \varphi\rangle\,.
\ee
The only configurations $\rho$ which contribute have $O$ in the first $n$ copies, that is $\rho=(\mathbb{1},\ldots,O,\ldots \mathbb{1},S)$ with $S$ containing only vacuum operators; all other configurations either vanish or lead to a disconnected piece. The sum over configurations $\rho$ thus turns into a sum of $j$ from $1$ to $n$, and the sum over $S_K$ is trivial since $K=0$. Assuming that $\check O$ is normalized to 1, we get $\kappa(N)= \frac1N$. Thus, (\ref{Iexpression}) turns into
\be
\frac1N \sum_{j,j'=1}^n \langle G_{-1/2} \varphi|O^{(j)}(1)O^{(j')}(x)|G_{-1/2} \varphi\rangle \, \label{appcorrc}.
\ee
By carrying out the steps laid out in Sec.\,\ref{sec:cfs} we end up with a linear combination of terms of the form of
\be
\sum_{j,j'=1}^n\frac{y_j^{s-\ell} {y'_{j'}}^{\ell}}{(y_j-y'_{j'})^s}\,,\label{eq:sumterm}
\ee
where $\ell\in\mathbb{Z}$, while $y$ and $y'$ are the inverse images under the covering map of 1 and $x$, respectively, such that
\be
y_j=\zeta_n^j\,,\quad\text{and}\quad y'_{j'}=x^{1/n}\zeta_n^{j'}\,.
\ee

In order to compute the sums in \eqref{appcorrc} we first rewrite \eqref{eq:sumterm} as
\be
\sum_{j,j'=1}^n\frac{\zeta_n^{sj-\ell(j-j')}x^{\ell/n}}{\left(\zeta_n^j-x^{1/n}\zeta_n^{j'}\right)^s}=\sum_{j,j'=1}^n\frac{\zeta_n^{(s-\ell)(j-j')}x^{\ell/n}}{\left(\zeta_n^{(j-j')}-x^{1/n}\right)^s}=n\sum_{j=1}^n\frac{\zeta_n^{j(s-\ell)}\alpha^{\ell}}{\left(\zeta_n^{j}-\alpha\right)^s}\,,
\ee
where we have defined $\alpha\coloneqq x^{1/n}$ and we have reduced the expression to just one sum over the $n$-th roots of unity. In order to compute the remaining sum we notice that
\be
z^n-1=\prod_{j=1}^n(z-\zeta_n^j)\,,
\ee
so $(z^n-1)^{-1}$ has simple poles located at $z=\zeta_n^j$, with residue
\be
\text{Res}_{z=\zeta_n^j}\left[\frac{1}{z^n-1}\right]=\lim_{z\rightarrow \zeta_n^j}\frac{z-\zeta_n^j}{z^n-1}=\lim_{z\rightarrow \zeta_n^j}\frac{1}{nz^{n-1}}=\frac{1}{n}\zeta_n^j\,.
\ee
Using this fact, we can define a function $f$ that has simple poles at the roots of unity with residue equal to precisely the terms that we want to sum over
\be
f(z)\coloneqq \frac{n^2 z^{s-\ell-1}\alpha^\ell}{(z-\alpha)^s}\frac{1}{z^n-1}\,.
\ee
One can easily check that this $f$ satisfies
\be
\text{Res}_{z=\zeta_n^j}\left[f(z)\right]=\frac{n\zeta_n^{j(s-\ell)}\alpha^\ell}{(\zeta_n^j-\alpha)^s}\,,
\ee
as claimed. Now, as long as $\ell\ge -n\,$, which is true for all the sums that we have to compute, $f(z)$ approaches zero as we increase $\abs{z}$. Therefore, it must be true that
\be
\lim_{r\rightarrow\infty}\oint_{C_{r,0}}dz \ f(z)=0\,,
\ee
and thus, by the residue theorem, it follows that the sum over all residues on the complex plane evaluates to zero. We thus conclude that
\be
n\sum_{j=1}^n\frac{\zeta_n^{j(s-\ell)}\alpha^{\ell}}{\left(\zeta_n^{j}-\alpha\right)^s}=\sum_{j=1}^n\text{Res}_{z=\zeta_n^j}\left[f(z)\right]=-\text{Res}_{z=\alpha}\left[f(z)\right]-\text{Res}_{z=0}\left[f(z)\right]\,.
\ee
Unpacking this equation leads to the following conclusion: Any sum that we encounter when computing \eqref{eq:basecfst} can be found by evaluating at most two residues.

\bibliographystyle{ytphys}
\bibliography{ref}

\end{document}